\documentclass[pra,twocolumn,floatfix,showpacs]{revtex4-1} 
\usepackage{graphicx,color}
\usepackage{amsmath,amssymb,bm}
\usepackage{comment} 
\usepackage{charter}
\usepackage[T1]{fontenc} 
\usepackage[colorlinks=true,linkcolor=blue,citecolor=red,urlcolor=magenta]{hyperref}

\begin{document} 

\title{Correlations in few two-component quantum walkers on a tilted lattice}

\author{Saubhik Sarkar}
\affiliation{\mbox{Institute of Physics, Polish Academy of Sciences, Aleja Lotnikow 32/46, PL-02668 Warsaw, Poland}}

\author{Tomasz Sowi\'{n}ski}
\affiliation{\mbox{Institute of Physics, Polish Academy of Sciences, Aleja Lotnikow 32/46, PL-02668 Warsaw, Poland}}

\begin{abstract}
We study the effect of inter-component interactions on the dynamical properties of quantum walkers. We consider the simplest situation of two indistinguishable non-interacting walkers on a tilted optical lattice interacting with a walker from a different component. The mediated effect of the third particle is then analyzed in the backdrop of various controlling parameters. The interaction-induced two-particle correlations are shown to be non-trivially affected by the particle statistics, choice of initial states, and tilt configurations of the lattice. Our analysis thus offers an overall picture and serves as a starting point of a study of interacting multi-component quantum walkers.    
\end{abstract}

\maketitle 
 
%%%%%%%%%%%%%%%%%%%%%%%%%%%%%%%%%%%%%%%%%%%%%%%%%%%%
\section{Introduction}
\label{Introduction}

Continuous-time quantum walk, where the particle moves on a lattice under the action of a time-independent Hamiltonian~\cite{Aharonov1993}, has attracted a lot of attention in recent years, especially due to its application in quantum information processing and quantum computation~\cite{Venegas-Andraca2012}. Recently it has also generated interest as a possible candidate as a quantum simulator of dynamics of magnon excitations in ferro- or antiferromagnetic solid-state systems~\cite{Fabiani2019,Wrzosek2020,Liu2019}. Experimental realizations of quantum walks are therefore of great importance and form the basis of interesting real-world applications. For a single walker, this has been realized in various setups, such as spin systems~\cite{Weitenberg2011,Fukuhara2013}, photonic systems~\cite{Manouchehri2014,Peruzzo2010}, trapped ions~\cite{Zahringer2010}, and with neutral cold atoms in optical lattices~\cite{Karski2009,Weitenberg2011,Fukuhara2013}. The last system is of pertinent interest as it provides a clean, coherent, and controllable way to investigate quantum many-body properties~\cite{Cirac2012,Bloch2012}. The optical lattices are nowadays a viable candidate for simulating condensed matter systems \cite{Jaksch2005,Cirac2012,Bloch2012} which have been instrumental in studies of strongly correlated many-body systems~\cite{Bloch2006,Bloch2008,Lewenstein2007,Lewenstein2017,Penna2017,Richaud2017,Keiler2018,Keiler2019}. Microscopic control of the Hamiltonian parameters via external fields in these setups allows precise probing of more interesting situations when instead of a single quantum walker, a larger number of particles is considered~\cite{Bromberg2009,Ahlbrecht2012}. It was argued that in the presence of inter-particle interactions, a many-particle quantum walk may affect quantum interference depending on the particle statistics~\cite{Qin2014} and therefore it also may have some applicability in universal quantum computation schemes~\cite{Childs2013}. This path of exploration is even more intriguing when the quantum walk is realised in the tilted lattice, {\it i.e.}, when local single-particle energies vary linearly with the site index. Starting from the famous theoretical paper by Bloch~\cite{1929BlochZeit} it is known that on the single-particle level the lattice tilt may lead to a counter-intuitive and quite spectacular effect of spatial oscillations of a single-particle density (known as Bloch  oscillations). As argued in~\cite{1929BlochZeit}, whenever transitions to higher bands are significantly suppressed by energy conservation arguments, action of the linear force results in oscillation of the single-particle quasi-momentum distribution in the first Brillouin zone. These oscillations are reflected in analogous oscillations of the particle wavepacket position. This prediction was awaiting for experimental confirmation over 60 years and was first demonstrated for electrons moving in a semiconductor superlattice \cite{1992LeoSolidStateCom}. Subsequent realizations in optical lattice experiments were first performed with non-interacting neutral atoms~\cite{Dahan1996,Wilkinson1996}. Quite recently, the Bloch oscillations phenomena has been used for precision measurements of force ~\cite{Ferrari2006,Poli2011,Gustavsson2008,Fattori2008} and in coherent matter wave transport~\cite{Alberti2009,Haller2010,Zhang2010}. The role of inter-particle interactions between indistinguishable walkers in the tilted lattice has been deeply explored in the literature \cite{Lyssenko1997,Hartmann2004,Dias2007,Khomeriki2010,Sachdev2002,Meinert2014,Preiss2015,Wiater2017,Mondal2020,Alonso_Lobo2018}. However, some questions still remain open and appropriate analysis is required. An important one of them is related to the problem of quantum walk realized in the tilted lattice by a multi-component system and to the non-obvious interplay between quantum statistics, initial state, tilt structure, as well as intra- and inter-component interactions. To make the first step to fill this gap, in our work we focus on the simplest multi-component system capturing all these features, {\it i.e.}, the case of three particles: two indistinguishable fermions or bosons belonging to the component $A$ and a third particle of fundamentally distinguishable flavor $B$. We study different dynamical properties of the system focusing mostly on quantum correlations between non-interacting $A$-particles induced by interactions with third $B$-walker. In this work we focus solely on the tilted lattice in which the dynamics is restricted to a finite spatial range. This dynamically induced confinement leads, in contrast to a ballistic spread in a uniform lattice case, to more pronounced effects of interactions and particle statistics.

This paper is organized in the following way. We provide the details of the system and the Hamiltonian in Sec.~\ref{System}. In Sec.~\ref{Two particles} we analyze the dynamics of two identical particles localized initially in distinct sites and highlight the differences forced by quantum statistics. Then, in Sec.~\ref{Three particles}, we introduce an additional particle of different component and study its impact on the dynamics. Here, we also look for the effects of particle statistics in the two-particle sector and uncover a role of the initial state. Next, in Sec.~\ref{Lattice} we show that inter-particle correlations may crucially depend on the tilt structure. Finally, we summarize and conclude in Sec.~\ref{Conclusion}. 

%%%%%%%%%%%%%%%%%%%%%%%%%%%%%%%%%%%%%%%%%%%%%%%%%%%%
\section{The system studied}
\label{System} 

In our work we consider the system of three quantum walkers experiencing the same one-dimensional tilted lattice potential. We assume that two of them are indistinguishable (bosons or fermions) and belong to the component $A$, while the third one belongs to the other component $B$. In the simplest case, when particles move in the lowest band of the lattice, the Hamiltonian of the system can be written in the tight-binding approximation as the following single-band Hubbard Hamiltonian~\cite{Hubbard1963,Jaksch1998}
\begin{align} 
\hat{ H} &= \displaystyle\sum_{i} \left[- J_{i} \left( \hat{a}^{\dag}_{i}\hat{a}_{i+1}+\hat{b}^{\dag}_{i}\hat{b}_{i+1} + h.c. \right) + E_{i}\left( \hat{n}_{A i} + \hat{n}_{B i} \right) \right] \nonumber \\ &+ U \sum_i \hat{n}_{A i}\hat{n}_{B i} + \frac{U_A}{2}\sum_i \hat{n}_{Ai} (\hat{n}_{Ai} -1),
\label{FH}
\end{align} 
where $J_{i}$ and $E_{i}$ characterize lattice geometry and determine single-particle tunnelings and on-site energies, respectively. Since we consider the simplest scenario of tilted lattice, in the following we set tunneling amplitudes $J_{i}=J$ as site-independent. On the contrary, the local energies, although component-independent, are linear in the site index, $E_{i}={ F} \times i$. Tilt parameter $F$ being under experimental control can be viewed as a constant force acting along the lattice. By definition, operators $\hat{a}_{i}$ and $\hat{b}_{i}$ annihilate particles at site $i$ belonging to the component $A$ and $B$, respectively. Depending on the quantum statistics, they obey intra-component commutation or anti-commutation relations. At the same time, any two operators acting in subspaces of different components do commute. Since in our work we consider systems containing only a single $B$-particle, all the results presented are insensitive to the intra-component quantum statistics of operators $\hat b_i$. For convenience, we introduced local number operators $\hat{n}_{A i} = \hat{a}^\dagger_i\hat{a}_i$ and $\hat{n}_{B i} = \hat{b}^\dagger_i\hat{b}_i$. 

In our work, we assume that the interaction part of the Hamiltonian is dominated by local terms. This assumption is particularly well justified for ultra-cold atomic systems interacting mainly via short-range intra- and inter-component interactions. Note, however, that in the case of fermionic $A$-particles, a double occupation at any site cannot occur due to the Pauli exclusion principle ($n_{Ai}\in\{0,1\}$ for any $i$). Thus, intra-component interaction terms controlled by $U_A$ rigorously vanish and only the inter-component terms (controlled by $U$) remain in play. In the case of bosonic $A$-particles additional intra-component interactions controlled by $U_A$ are possible. In most of the cases studied here, we are interested in the case of non-interacting $A$-particles ($U_A=0$). In view of these remarks, the Hamiltonian~\eqref{FH} is appropriately written for bosonic as well as for fermionic particles.

%%%%%%%%%%%%%%%%%%%%%%%%%%%%%%%%%%%%%%%%%%%%%%%%%%%%
\section{Dynamics of two particles} 
\label{Two particles}

\begin{figure}[t]
\centering
\includegraphics[width=\linewidth]{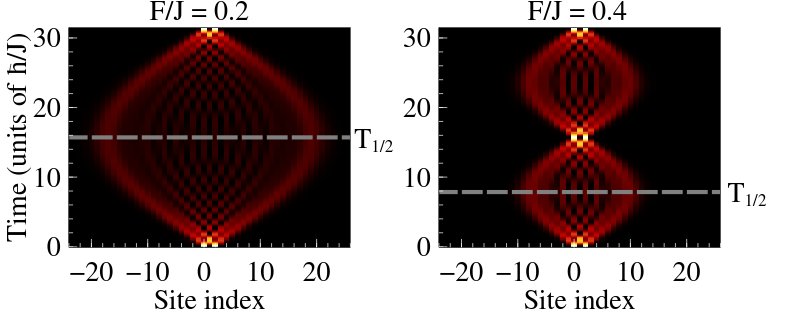}
\label{figure}\caption{Evolution of the single-particle density profile $\gamma_{Ai}$ for two different tilt values of the lattice and initial distance between occupied sites $d=2$. Horizontal dashed lines indicate half-time of Bloch oscillation period $T_{1/2}$. As argued in the main text, these results are independent of the quantum statistics of $A$-particles. Time is measured in the natural unit of the problem $\hbar/J$.\label{Fig1} }
\end{figure}

To make further analysis clearer, let us first make some observations on the dynamics of two identical particles belonging to the component $A$ initially occupying two different sites near the center of the lattice, {\it i.e.}, as the initial state we take $|\Psi_0\rangle = \hat{a}^\dagger_{0}\hat{a}^\dagger_{d}|\mathtt{vac}\rangle$, where $d\neq 0$ denotes the distance between occupied sites. In this case the Hamiltonian \eqref{FH} is reduced to the single component and has the following form
\begin{align} 
\hat{H} &= \sum_{i}\left[ - { J}\left(\hat{a}^{\dag}_{i}\hat{a}_{i+1}+\hat{a}^{\dag}_{i+1}\hat{a}_{i}\right) + { F}\,i\, \hat{n}_{Ai} \right].
\end{align} 
It is very instructive to note that for the initial state considered here, the time evolution of a whole single-particle density matrix 
\begin{equation}
\rho_{ij}(t) = \langle\Psi(t)|\hat{a}^\dagger_{i}\hat{a}_{j}|\Psi(t) \rangle,
\end{equation}
where $|\Psi(t)\rangle = \mathrm{e}^{-i\hat{H}t/\hbar}|\Psi_0\rangle$, 
is independent of the statistics, {\it i.e.}, it is identical for non-interacting bosons and fermions. Indeed, the commutation relation 
\begin{align}
\left[\hat{H}, \hat{a}^{\dag}_{i} \hat{a}_{j}\right] &= { J} \left[\hat{a}^{\dag}_{i}( \hat{a}_{j+1} + \hat{a}_{j-1}) - (\hat{a}^{\dag}_{i+1} + \hat{a}^{\dag}_{i-1}) \hat{a}_{j}\right] \nonumber \\
& + { F} (i-j) \hat{a}^{\dag}_{i} \hat{a}_{j}
\end{align}
is same for both non-interacting particles. The expectation values of this commutator and all the higher order commutators with the Hamiltonian in the localized initial state $|\Psi_0\rangle$ are insensitive to the particle statistics. Therefore, the single-particle density matrix $\rho_{ij}(t)$ evolves identically for bosons as well as for fermions. This means that any single-particle measurement is not able to capture any dynamical difference between both statistics if at initial moment particles do not occupy the same site. This result is quite counterintuitive since it is clear that tunneling to an already occupied neighboring site is blocked only for fermions. In Fig.~\ref{Fig1} we display the time evolution of the single-particle density profile $\gamma_{Ai}(t)=\langle\Psi(t)|\hat{n}_{Ai}|\Psi(t) \rangle$ (diagonal part of $\rho_{ij}(t)$) for two different tilt values of the lattice and initial distance $d=2$. Clearly visible characteristic spatial oscillations are exactly the same for bosons and fermions and their frequency depends only on the tilt strength. Half of the oscillation period is given by $T_{1/2} = \pi \hbar/F$, and this is when the particles reach the farthest part of the lattice. Note that this maximal distance from the initial position decreases with increasing tilt strength, which is in full agreement with previously obtained results~\cite{Hartmann2004}.

\begin{figure}[t]
\centering
  \includegraphics[width=\linewidth]{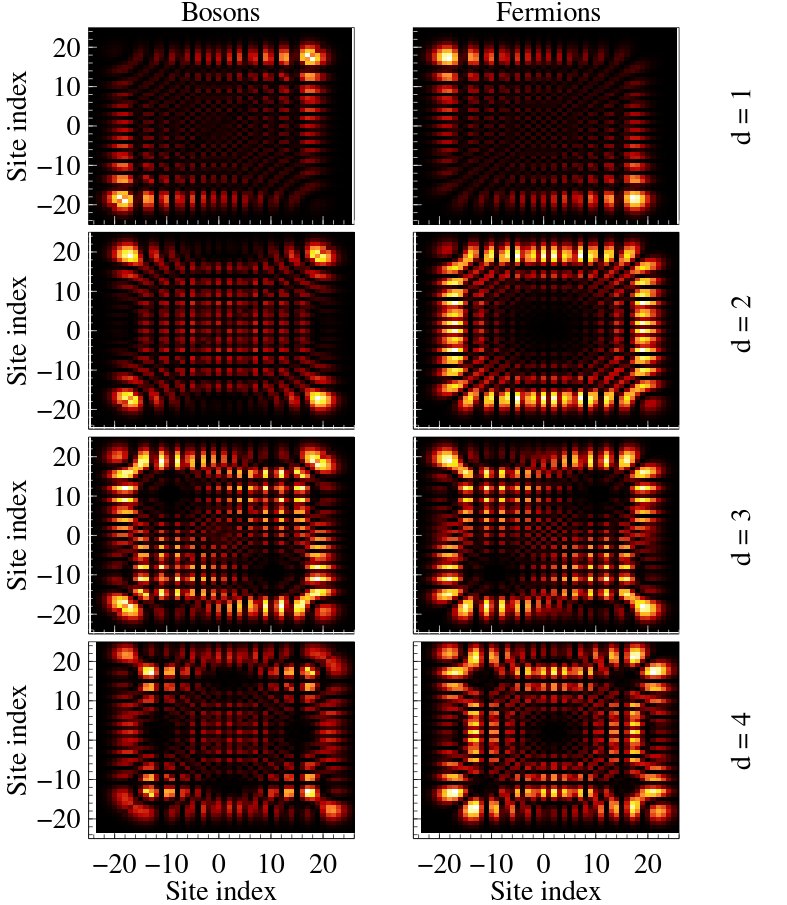}
\label{figure}\caption{Two-particle density profile $\Gamma_{ij}$ at the half-time of a Bloch oscillation period $T_{1/2}$ with tilt value $F/J = 0.2$ for a system of two non-interacting bosons (left) and fermions (right) initially prepared in the state $|\Psi_0\rangle = \hat{a}^\dagger_{0}\hat{a}^\dagger_{d}|\mathtt{vac}\rangle$. Note that, regardless of the initial distance $d$ the distributions are significantly distinct for different statistics. \label{Fig2} }
\end{figure}

The dynamical difference between bosonic and fermionic systems is however well-captured by higher-order correlations. In the case of two-particle systems one of them is particularly important since it can be obtained directly by repeated instant-time measurements of particles' positions -- the two-particle density profile
\begin{equation} \label{2bodyprofile}
\Gamma_{ij}(t) = \langle\Psi(t)|\hat{n}_{Ai}\hat{n}_{Aj}|\Psi(t)\rangle.
\end{equation}
In this case the commutation with the Hamiltonian would again have an identical operator form for bosons and fermions, namely,
\begin{align}
\left[\hat{H},\hat{n}_{Ai}\hat{n}_{Aj}\right]  
&= {J} \! \left[\hat{a}^{\dag}_{i}( \hat{a}_{i+1} \! + \! \hat{a}_{i-1}) \! - \! (\hat{a}^{\dag}_{i+1} \! + \! \hat{a}^{\dag}_{i-1}) \hat{a}_{i}\right] \! \hat{n}_{Aj} \nonumber \\
&+ {J} \hat{n}_{Ai} \left[\hat{a}^{\dag}_{j}( \hat{a}_{j+1} + \hat{a}_{j-1}) \! - \! (\hat{a}^{\dag}_{j+1} + \hat{a}^{\dag}_{j-1}) \hat{a}_{j}\right].
\end{align}
Subsequently, the evaluation of $\Gamma_{ij}(t)$ requires determination of expectation values of this commutator (and its higher orders) consisting different two-particle operators generically having the following form
\begin{align}
\hat{a}^\dagger_{i} \hat{a}_{i'} \hat{a}^\dagger_{j} \hat{a}_{j'}  =  \hat{a}^\dagger_{i} \hat{a}_{j'} \delta_{i'j} \pm \hat{a}^\dagger_{i} \hat{a}^\dagger_{j} \hat{a}_{i'} \hat{a}_{j'},
\end{align}
where top and bottom signs are used for bosons and fermions, respectively. It implies that whenever $\Gamma_{ij}$ is evaluated, normal-ordered two-particle operators will typically generate different contributions in the localized initial state for different statistics. This can be checked straightforwardly, for example, by simply looking at the case $i=j'=0, i'=j=d$. As a consequence, the total initial expectation values are not the same any more. It is also clear at this point that for different initial states, {\it i.e.}, for different distances $d$, different higher-order terms will interfere constructively or destructively and thus will produce a specific $d$-dependent correlation pattern. The pattern can be treated therefore as a peculiar fingerprint of the initial distance $d$ and the quantum statistics. These  different dynamical behaviors are clearly visible when two-particle profiles are compared at the half-time of a Bloch oscillation period, $T_{1/2}$ (see Fig.~\ref{Fig2}). The effect of particle statistics is also clear as, independent of the initial distance $d$, two-particle densities are essentially different not only close to the diagonal of the doublon occupancy which is forbidden for fermions, but also in a whole spatial range accessible to particles. At this point it is worth to note that such a two-particle correlation dynamics has been extensively studied for various other aspects in literature. For example, in~\cite{Qin2014} the effects of particle statistics and next-neighbor interactions were studied in absence of tilt, while the hard-core boson counterparts with $d=1$ in tilted lattice were studied recently in~\cite{Liu2019}.

\begin{figure}[t]
\centering
\includegraphics[width=0.8\linewidth]{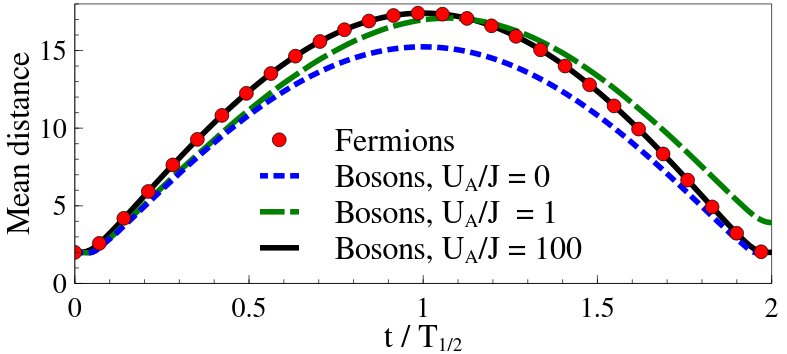}
\label{figure}\caption{Time dependence of mean distance $D(t)$ between non-interacting $A$-particles prepared in the state $|\Psi_0\rangle = \hat{a}^\dagger_{0}\hat{a}^\dagger_{d}|\mathtt{vac}\rangle$ with $d=2$ in presence of tilt strength $F/J = 0.2$. Due to the bosonic enhancement, bosons spread slower than fermions and only in the limit of hard-core interactions ($U_A/J\rightarrow\infty$) both statistics are indistinguishable by this quantity. Note that, for finite non-vanishing interactions between particles the Bloch oscillations are slightly slowed down and the system does not return exactly to the initial state after the Bloch period. \label{Fig3} }
\end{figure}

This significant difference between bosonic and fermionic particles can be quantified with another quantity that is important from a measurement point of view -- the mean distance between particles, $D(t) =\sum_{ij}|i-j| \, \Gamma_{ij}(t)$. The bosonic enhancement results in a smaller mean distance between non-interacting bosons when compared to the fermionic case. In Fig.~\ref{Fig3} we present the mean distance $D(t)$ for the particular initial state with $d=2$. For a complete picture, we display also the bosonic distance $D(t)$ when the on-site repulsion term in the Hamiltonian controlled by $U_A$ is present. For smaller values of $U_A$, the inter-particle interaction dephases the Bloch oscillations~\cite{Buchleitner2003} and the time at which the bosons are maximally separated starts moving to the right. However, this time of maximum separation starts decreasing again for stronger values of $U_A$, as the dynamics starts to be dominated by a single-particle physics again. It is clear, that as the interaction increases the mean distance for the bosons approaches the fermionic case. In the limiting scenario of hard-core bosons (${ U}_A/{ J}\rightarrow\infty$) both statistics display the same behavior.

%%%%%%%%%%%%%%%%%%%%%%%%%%%%%%%%%%%%%%%%%%%%%%%%%%%%
\section{Influence of the third particle}
\label{Three particles}

After discussing the dynamical consequences of statistics for indistinguishable particles we focus on a system of three particles. In the following, we assume that two particles belonging to the $A$-component (bosons or fermions) do not interact with each other but their dynamics is affected by a third $B$-particle via on-site interaction term of the form $U \sum_i \hat{n}_{A i}\,\hat{n}_{B i}$. In this way, we want to figure out how the dynamical features of two-particle correlations described above are affected by the presence of additional, fundamentally different, particle.

The role played by the third particle crucially depends not only on interaction strength $U$ but also on its initial state. Therefore, we consider a whole family of initial states of the form of Gaussian wave packets localized exactly between $A$-particles. The initial configuration of the two $A$-particles is same as in the previous case. All it means is that the initial state of the whole system reads
\begin{equation} \label{Initial_State}
|\Psi_0 \rangle ={\cal N}\, \hat{a}^{\dag}_{0}\hat{a}^{\dag}_{d}\left(\sum_i e^{-|i-d/2|^2/2 \sigma^2} \hat{b}^{\dag}_{i}\right) |\mathtt{vac}\rangle,
\end{equation}
where $\cal N$ is the normalization factor and $\sigma$ determines width of the component $B$ wave packet. The limit of ideally localized $B$-particle ($\sigma=0$) is well-defined only for even $d$ ($B$-particle is localized exactly at site $i=d/2$). Therefore, to capture also this limiting case, we focus mainly on the $d=2$ case in the following.

\begin{figure}[t]
\centering
\includegraphics[width=\linewidth]{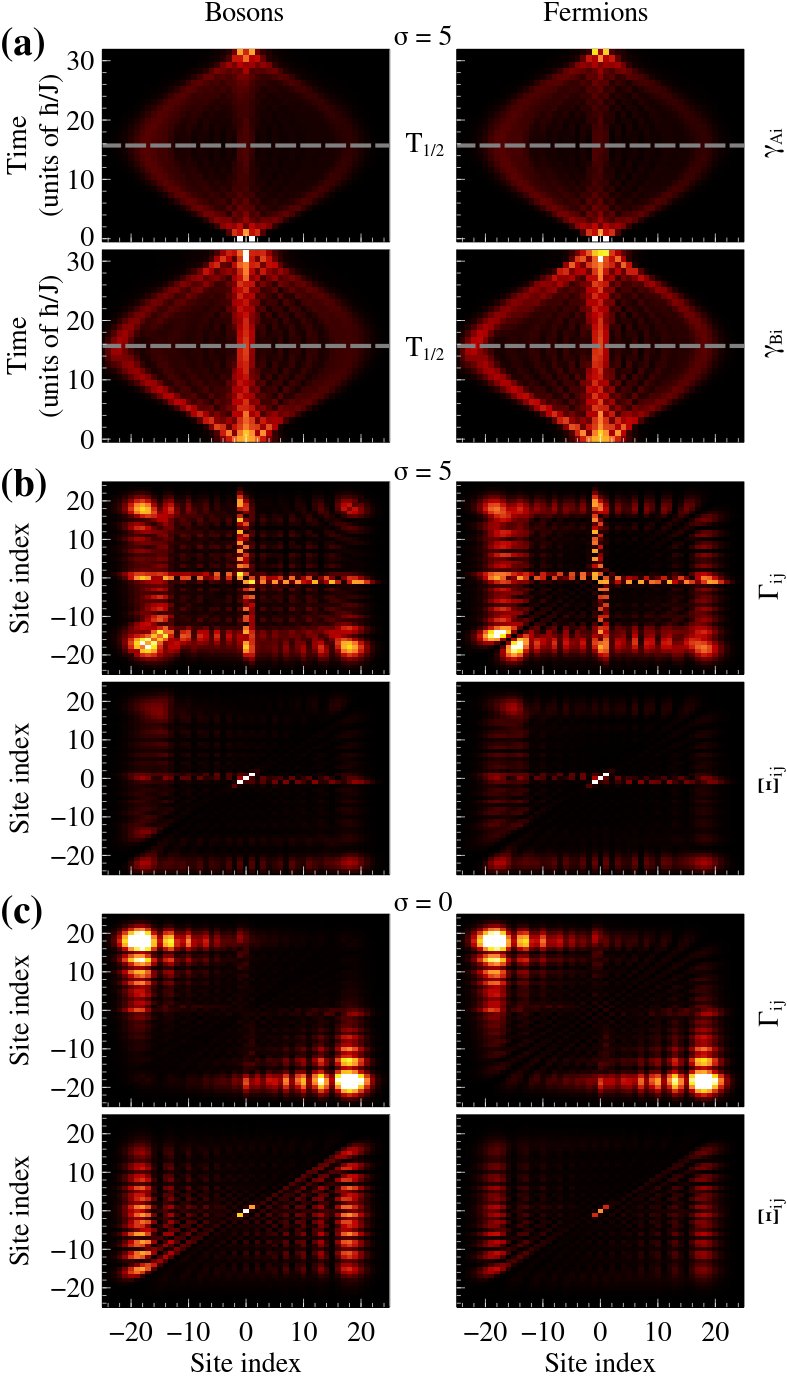} 
\caption{(a) Evolution of the single-particle density profiles for a system of two bosonic (left) and fermionic (right) $A$-particles initially localized at distant $d=2$ and interacting with $B$-particle ($U/J=10$) being delocalized with $\sigma=5$. (b) Corresponding two-particle density profile $\Gamma_{ij}$ of $A$-particle and the inter-component correlation $\Xi_{ij}$ at the half-time of a Bloch oscillation period $T_{1/2}$ for the same conditions. (c) For contrast, $\Gamma_{ij}$ and $\Xi_{ij}$ at $T_{1/2}$ when the $B$-particle is initially perfectly localized ($\sigma=0$). Note that, for delocalized $B$-particle a specific `cross-like' structure in $\Gamma_{ij}$ is enhanced together with appearance of partial pairing. Compare with Fig.~\ref{Fig5} for other initial distances $d$. In all plots tilt strength $F/J = 0.2$. \label{Fig4} }
\end{figure}

At this point, it should be noted that in absence of interactions, dynamics of the Gaussian wave packet is reduced to simple oscillations without change of the shape, provided that width $\sigma$ is larger than the periodicity of the lattice~\cite{Hartmann2004,Wiater2017}. On the contrary, when inter-component interactions are present ($U \neq 0$), any non-zero $\sigma$ introduces finite initial interaction energy to the system. 

Exactly as in the case of two particles, we inspect the dynamics of the system during one Bloch oscillation (Fig.~\ref{Fig4}). Firstly, differences caused by the $B$-particle are visible already in the single-particle distribution of the $A$-component $\gamma_{Ai}(t)$ and the analogous quantity $\gamma_{Bi}(t)=\langle\Psi(t)|\hat{n}_{Bi}|\Psi(t) \rangle$ for $B$-particle (Fig.~\ref{Fig4}a). However, distinctions between fermionic and bosonic $A$-particles are still not substantial. Therefore, we capture the differences caused by interactions and quantum statistics at the moment when they become the largest, {\it i.e.}, at the half-time of the Bloch oscillation period $T_{1/2}$ with the tilt strength fixed at $F/J=0.2$. The two-particle density profile $\Gamma_{ij}(t)$ and the inter-component density correlation $\Xi_{ij}(t)$ defined as
\begin{equation}
\Xi_{ij}(t)=\langle\Psi(t)|\hat{n}_{Ai}\hat{n}_{Bj}|\Psi(t)\rangle
\end{equation}
at $T_{1/2}$ in the case of strong inter-component interactions $U/J=10$ and delocalization $\sigma=5$ are shown in Fig.~\ref{Fig4}b. We compare this with $\Gamma_{ij}(t)$ and $\Xi_{ij}(t)$ at $T_{1/2}$ computed for the case when the $B$-particle is perfectly localized ($\sigma=0$) in Fig.~\ref{Fig4}c.

Depending on the quantum statistics of $A$-particles and the width of Gaussian packet $\sigma$ different correlations are enhanced. When the $B$-particle is precisely localized ($\sigma=0$, Fig.~\ref{Fig4}c), there is no significant difference between bosonic and fermionic particles. It means that strong interaction with the $B$-particle localized exactly between the $A$-particles changes the behavior considerably, even in the absence of initial interaction energy. Significant differences caused by the statistics, which was observed previously (compare with Fig.~\ref{Fig2}), are almost completely blurred. In both cases, the probability of finding particles on opposite sides of the central site is the largest. On top of this clear correlation, one finds also a very weak `cross-like' structure emerging in the density profile. It uncovers additional enhancement of probability of finding one particle near the center site with the second particle smeared over a whole available space.  

\begin{figure}[t]
\centering
\includegraphics[width=\linewidth]{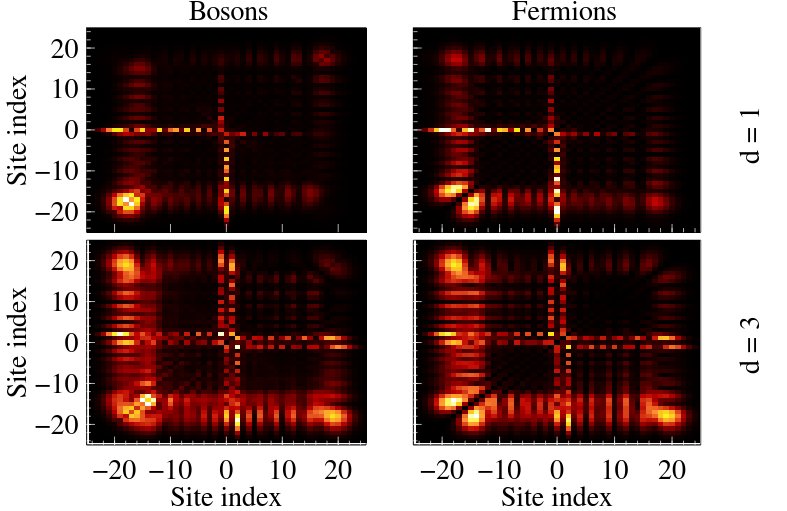}
\caption{Two-particle density profile $\Gamma_{ij}$ at the half-time of a Bloch oscillation period $T_{1/2}$ with tilt strength $F/J = 0.2$ for a system of two bosons (left) and fermions (right) interacting ($U/J=10$) with the $B$-particle initially delocalized with $\sigma=5$. The initial distance between localized $A$-particles is $d=1$ and $d=3$ (top and bottom rows, respectively). Compare with Fig.~\ref{Fig4} for other initial delocalization $\sigma$ of the $B$-particle.\label{Fig5} }
\end{figure}

The situation is significantly different when the $B$-particle is initially delocalized over several sites ($\sigma =5$, Fig.~\ref{Fig4}b). For both statistics of $A$-particles we observe a noticeable enhancement of the `cross-like' structure in the distribution. Moreover, the inter-component repulsion causing strong confinement of the inter-component density correlation $\Xi_{ij}(t)$, supports the pair-like formation of $A$-particles towards the left edge of the lattice (direction favored by the tilt of the lattice). We checked that this effect remains the same if the sign of inter-component interactions is changed to attractions. It is also independent of the initial distance between particles $d$ provided that the width $\sigma$ is large enough to make mediation between $A$-particles possible (see Fig.~\ref{Fig5}). This result evidently shows that a specific attraction between non-interacting particles is dynamically induced when particles do interact with other particle which necessarily needs to be sufficiently delocalized.

\begin{figure}[t]
\centering
\includegraphics[width=\linewidth]{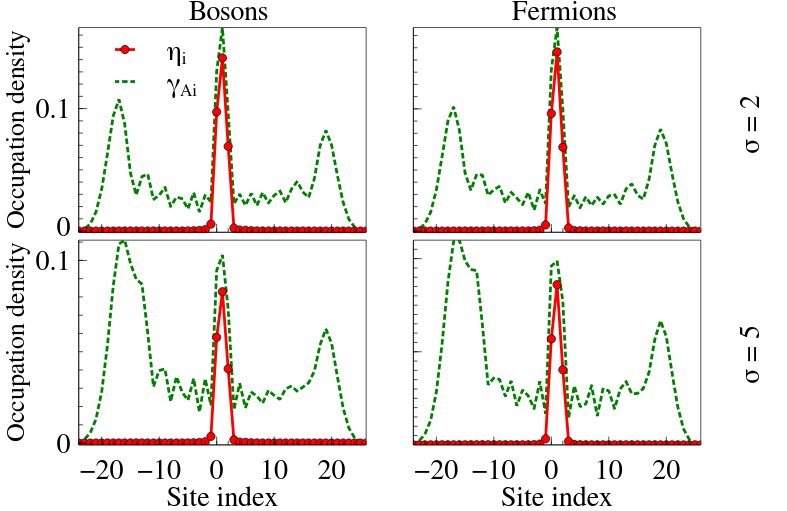}
\label{figure}\caption{Density distributions for the strongly interacting system ($U/J=10$) initially prepared in the state \eqref{Initial_State} with $d=2$ and $\sigma=2$ (top row), $\sigma=5$ (bottom row) and at the half-time of Bloch period $T_{1/2}$ with tilt strength $F/J = 0.2$. Dashed green line and red solid lines correspond to the single-particle density of $A$-particles $\gamma_{Ai}$ and the doublon density $\eta_i$, respectively. \label{Fig6} }
\end{figure}

To get a better understanding of these non-trivial correlations appearing in the system, besides computing the two-particle density profiles $\Gamma_{ij}$, we also consider the density profile of doublons $\eta_i(t)$. This is given by the probability density of finding at least one $A$-particle and $B$-particle together at the same site. In the case of fermionic particles it can be calculated straightforwardly as $\eta_i(t) = \langle \Psi(t)|\hat{n}_{Ai} \hat{n}_{Bi}|\Psi(t) \rangle$, {\it i.e.}, it is the diagonal part of the inter-component density $\Xi_{ij}(t)$. For bosons, due to possible double-occupancy of $A$-particles, the definition is slightly modified and in the case studied can be expressed as $\eta_i(t) = \langle\Psi(t)|n_{Ai} n_{Bi} (3 - n_{Ai})/2 |\Psi(t)\rangle$. It is important to note, that due to a well-known halving of the time period of the Bloch oscillation for doublons~\cite{Dias2007,Khomeriki2010}, the doublon density is particularly very straightforward at time $T_{1/2}$. Exactly at this moment, when the single-particle density peaks are maximally distant from the origin, the doublon density $\eta_i$ is non-zero only in the vicinity of the central site. In Fig.~\ref{Fig6} we illustrate this effect for $d=2$ and different initial widths $\sigma$ of the $B$-particle wave packet and different statistics of $A$-particles (solid red lines for doublon density $\eta_i$ and dashed green lines for the single-particle density $\gamma_{Ai}$). This specific composition of the doublon density and the single-particle profile results in the final two-particle correlation of $A$-particles. The `cross-like' structure is triggered mostly by the doublon, while the remaining part at the edges of the system comes from the distribution of the remaining $A$-particle. The height of the peak in the central region, for the doublon density $\eta_i(t)$ as well as for the single-particle density $\gamma_{Ai}(t)$, is dependent on the initial overlap of $A$- and $B$-particle distributions, which is the highest at around $\sigma=2$ for this particular initial state ($d=2$). Therefore, for larger values of $\sigma$ these peaks get diminished.

%%%%%%%%%%%%%%%%%%%%%%%%%%%%%%%%%%%%%%%%%%%%%%%%%%%%
\section{Role of the tilt structure}
\label{Lattice}

As we have shown, the two-body correlations between non-interacting $A$-particles forced by the presence of the additional particle depends crucially on the initial state. Interestingly and quite counterintuitively these correlations are less sensitive to the initial distance between particles and their statistics provided that inter-component interactions are strong enough. For completeness, we will now focus on the role of tilt structure in these correlations. One possible way of this occurring during a realization can come from inhomogeneity in the slope of the linear potential arising from experimental imperfections. To provide a concrete example, let us consider a one-dimensional lattice that is not homogeneously tilted but the tilt becomes inverted around the center of the initial state, {\it i.e.}, a lattice having local energies of the form $E_i = -F |i-i_0|$. Such an exotic tilt, although simple for theoretical consideration, may be demanding for experimental manufacturing. However, in the view of recent progress in creating very different lattice configurations, it is not impossible. One particular way of achieving a change of the slope could be engineered by initiating the particles at the center of a symmetric confinement. Then the dynamics can studied after quenching the trap to an inhomogeneous double-well setting where the slope changes strongly near the center. Another possibility is to use the standard magnetic field gradient~\cite{Preiss2015}, but with additional microscopic spatial dependence of the field that have been used in the past for imaging~\cite{Wildermuth2005}.  At this point it worth to point out that our arguments on the dynamics of two non-interacting particles presented in Sec.~\ref{Two particles} remain valid: at any moment, any single-particle quantity persists insensitive to the quantum statistics provided that initially particles are localized in two different sites of the lattice.

\begin{figure}[t] 
\centering
\includegraphics[width=\linewidth]{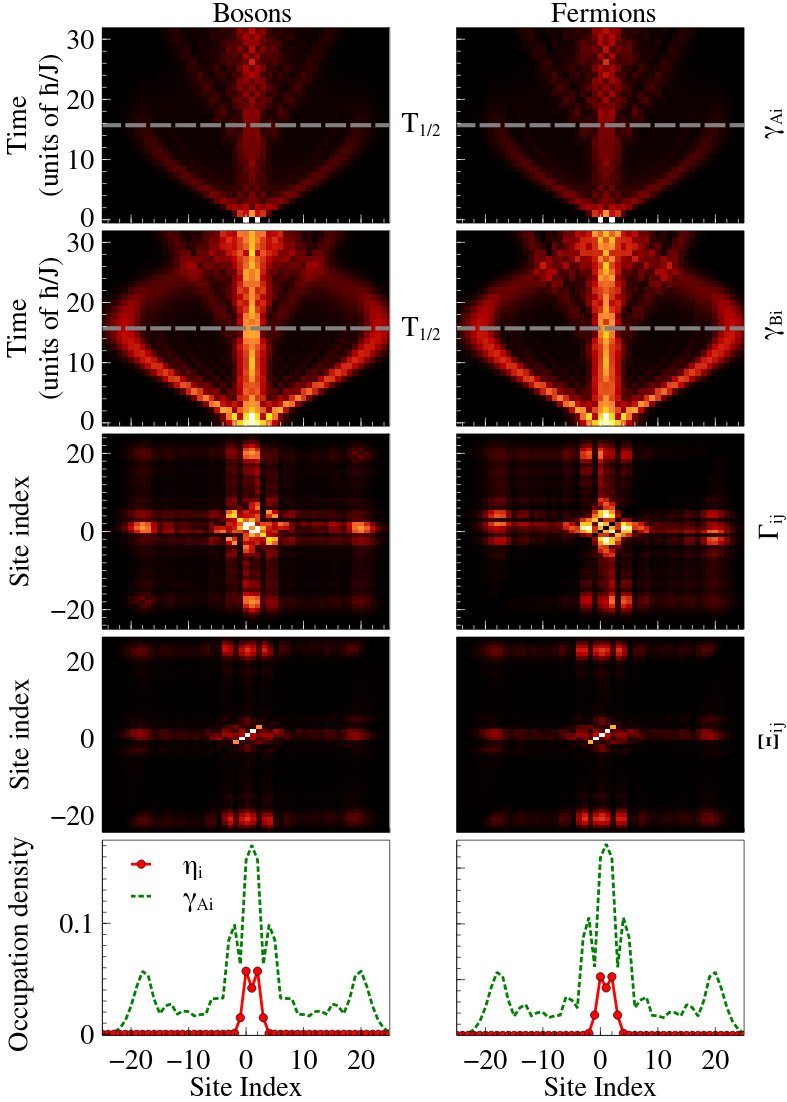}
\label{figure}\caption{Dynamics for a lattice that is tilted in opposite directions ($E_i = -F |i-i_0|$ with $F/J=0.2$), with two $A$-particles localized on both sides of the center ($d=2$), interacting with the $B$-particle centered around the middle ($U/J = 10$, $\sigma = 5$). Evolution of the single-particle density profiles $\gamma_{Ai}$ and $\gamma_{Bi}$ are shown in the first two rows. Next three rows show respectively: the two-particle density profile $\Gamma_{ij}$, the inter-component correlations $\Xi_{ij}$, and the doublon density $\eta_i$ (red line) with the single-particle density $\gamma_{Ai}$ (dashed green line), at the characteristic time $T_{1/2}$. \label{Fig7} }
\end{figure}

As the simplest example, we focus on the three-particle initial state \eqref{Initial_State} with $d=2$ and $\sigma=5$ centered exactly around the middle point of the lattice ($i_0=1$) with inter-component interaction $U/J=10$ and tilt strength $F/J = 0.2$. The dynamical situation of such a configuration is presented in Fig.~\ref{Fig7}. Exactly as previously, columns correspond to different statistics of $A$-particles. The single-particle density profiles, respectively for $A$ and $B$ particles, are displayed in the first two rows. The third row presents the two-particle density profile $\Gamma_{ij}$ at the half-time of a Bloch oscillation period $T_{1/2}$. It is clear that, exactly as in the previous case, all the features of the system are almost the same for both statistics. However, the two-particle correlation is strongly influenced by the change of the lattice configuration -- breaking of the lattice supports the `cross-like' structure with a magnification of density exactly in the center of the system. This increased concentration of $A$-particles in the center is also clearly visible in the single-particle density profile (bottom row, dashed green line). Note that in this case it is not triggered by the existence of doublons which is highly suppressed (red line) but confined in the center, as is the inter-component density correlation $\Xi_{ij}(t)$. As the two-particle correlations are substantially different from the previous case (see Figs.~\ref{Fig4} and~\ref{Fig5}) we systematically change the tilt configuration from the uniform structure (considered in the preceding sections) to the `broken' configuration (reported in this section) and compute the correlations. As the correlation pattern changes, we see the doublon peak gradually diminishing at the center of the lattice, where the single-particle density consequently increases. Although not displayed here, this behavior is confirmed in our calculations. 

%%%%%%%%%%%%%%%%%%%%%%%%%%%%%%%%%%%%%%%%%%%%%%%%%%%%
\section{Conclusion}
\label{Conclusion}

We have analyzed the dynamics of the simplest system of two-component quantum walkers on a tilted optical lattice and studied the effect of particle statistics, interaction, choice of initial states, and tilt structure, which lead to three main findings. Firstly, single-particle measurements cannot distinguish the quantum statistical nature of two non-interacting walkers during evolution from initially localized states. One must perform two-particle measurements to differentiate between the particle-statistics. Secondly, interaction with the third walker from a different distinguishable component can induce non-trivial correlations between two walkers which can be controlled by changing the initial states. As our third and final result, we show that the two-particle correlations can also be significantly modified by changing the tilting structure of the lattice. Implementation of a non-trivial tilt is a challenging task, however, with the ever-growing standard of ongoing optical lattice experiments with a controlled number of cold atoms, all the results presented in this work can be realized and bench-marked. Although the three-particle results presented in this study are numerical in nature, extensions can be made in future to gain some analytical insights within pertubative approach in the strong interaction limit. Furthermore, this study can be expanded to probe the effect of intra-component interactions, next-neighbor interactions (both inter- and intra-component in nature), single-particle state preparations, and a generalized study of the dynamics of multi-component quantum walkers, to name a few.   

%%%%%%%%%%%%%%%%%%%%%%%%%%%%%%%%%%%%%%%%%%%%%%%%%%%%
\begin{acknowledgements}
This work was supported by the (Polish) National Science Center Grant No.~2016/22/E/ST2/00555. The computational works were performed using the Interdisciplinary Centre for Mathematical and Computational Modelling, University of Warsaw (ICM), under Computational Grant No.~G75-6.
\end{acknowledgements}

%%%%%%%%%%%%%%%%%%%%%%%%%%%%%%%%%%%%%%%%%%%%%%%%%%%%
\bibliographystyle{apsrev4-1}
\bibliography{manuscript}

%merlin.mbs apsrev4-1.bst 2010-07-25 4.21a (PWD, AO, DPC) hacked
%Control: key (0)
%Control: author (72) initials jnrlst
%Control: editor formatted (1) identically to author
%Control: production of article title (-1) disabled
%Control: page (0) single
%Control: year (1) truncated
%Control: production of eprint (0) enabled
\begin{thebibliography}{51}%
\makeatletter
\providecommand \@ifxundefined [1]{%
 \@ifx{#1\undefined}
}%
\providecommand \@ifnum [1]{%
 \ifnum #1\expandafter \@firstoftwo
 \else \expandafter \@secondoftwo
 \fi
}%
\providecommand \@ifx [1]{%
 \ifx #1\expandafter \@firstoftwo
 \else \expandafter \@secondoftwo
 \fi
}%
\providecommand \natexlab [1]{#1}%
\providecommand \enquote  [1]{``#1''}%
\providecommand \bibnamefont  [1]{#1}%
\providecommand \bibfnamefont [1]{#1}%
\providecommand \citenamefont [1]{#1}%
\providecommand \href@noop [0]{\@secondoftwo}%
\providecommand \href [0]{\begingroup \@sanitize@url \@href}%
\providecommand \@href[1]{\@@startlink{#1}\@@href}%
\providecommand \@@href[1]{\endgroup#1\@@endlink}%
\providecommand \@sanitize@url [0]{\catcode `\\12\catcode `\$12\catcode
  `\&12\catcode `\#12\catcode `\^12\catcode `\_12\catcode `\%12\relax}%
\providecommand \@@startlink[1]{}%
\providecommand \@@endlink[0]{}%
\providecommand \url  [0]{\begingroup\@sanitize@url \@url }%
\providecommand \@url [1]{\endgroup\@href {#1}{\urlprefix }}%
\providecommand \urlprefix  [0]{URL }%
\providecommand \Eprint [0]{\href }%
\providecommand \doibase [0]{http://dx.doi.org/}%
\providecommand \selectlanguage [0]{\@gobble}%
\providecommand \bibinfo  [0]{\@secondoftwo}%
\providecommand \bibfield  [0]{\@secondoftwo}%
\providecommand \translation [1]{[#1]}%
\providecommand \BibitemOpen [0]{}%
\providecommand \bibitemStop [0]{}%
\providecommand \bibitemNoStop [0]{.\EOS\space}%
\providecommand \EOS [0]{\spacefactor3000\relax}%
\providecommand \BibitemShut  [1]{\csname bibitem#1\endcsname}%
\let\auto@bib@innerbib\@empty
%</preamble>
\bibitem [{\citenamefont {Aharonov}\ \emph {et~al.}(1993)\citenamefont
  {Aharonov}, \citenamefont {Davidovich},\ and\ \citenamefont
  {Zagury}}]{Aharonov1993}%
  \BibitemOpen
  \bibfield  {author} {\bibinfo {author} {\bibfnamefont {Y.}~\bibnamefont
  {Aharonov}}, \bibinfo {author} {\bibfnamefont {L.}~\bibnamefont
  {Davidovich}}, \ and\ \bibinfo {author} {\bibfnamefont {N.}~\bibnamefont
  {Zagury}},\ }\href {\doibase 10.1103/PhysRevA.48.1687} {\bibfield  {journal}
  {\bibinfo  {journal} {Phys. Rev. A}\ }\textbf {\bibinfo {volume} {48}},\
  \bibinfo {pages} {1687} (\bibinfo {year} {1993})}\BibitemShut {NoStop}%
\bibitem [{\citenamefont {Venegas-Andraca}(2012)}]{Venegas-Andraca2012}%
  \BibitemOpen
  \bibfield  {author} {\bibinfo {author} {\bibfnamefont {S.~E.}\ \bibnamefont
  {Venegas-Andraca}},\ }\href {\doibase 10.1007/s11128-012-0432-5} {\bibfield
  {journal} {\bibinfo  {journal} {Quantum Information Processing}\ }\textbf
  {\bibinfo {volume} {11}},\ \bibinfo {pages} {1015} (\bibinfo {year}
  {2012})}\BibitemShut {NoStop}%
\bibitem [{\citenamefont {{Fabiani}}\ and\ \citenamefont
  {{Mentink}}(2019)}]{Fabiani2019}%
  \BibitemOpen
  \bibfield  {author} {\bibinfo {author} {\bibfnamefont {G.}~\bibnamefont
  {{Fabiani}}}\ and\ \bibinfo {author} {\bibfnamefont {J.~H.}\ \bibnamefont
  {{Mentink}}},\ }\href {https://arxiv.org/abs/1912.10845} {\bibfield
  {journal} {\bibinfo  {journal} {arXiv:1912.10845}\ } (\bibinfo {year}
  {2019})}\BibitemShut {NoStop}%
\bibitem [{\citenamefont {Wrzosek}\ \emph {et~al.}(2020)\citenamefont
  {Wrzosek}, \citenamefont {Wohlfeld}, \citenamefont {Hofmann}, \citenamefont
  {Sowi\ifmmode~\acute{n}\else \'{n}\fi{}ski},\ and\ \citenamefont
  {Sentef}}]{Wrzosek2020}%
  \BibitemOpen
  \bibfield  {author} {\bibinfo {author} {\bibfnamefont {P.}~\bibnamefont
  {Wrzosek}}, \bibinfo {author} {\bibfnamefont {K.}~\bibnamefont {Wohlfeld}},
  \bibinfo {author} {\bibfnamefont {D.}~\bibnamefont {Hofmann}}, \bibinfo
  {author} {\bibfnamefont {T.}~\bibnamefont {Sowi\ifmmode~\acute{n}\else
  \'{n}\fi{}ski}}, \ and\ \bibinfo {author} {\bibfnamefont {M.~A.}\
  \bibnamefont {Sentef}},\ }\href {\doibase 10.1103/PhysRevB.102.024440}
  {\bibfield  {journal} {\bibinfo  {journal} {Phys. Rev. B}\ }\textbf {\bibinfo
  {volume} {102}},\ \bibinfo {pages} {024440} (\bibinfo {year}
  {2020})}\BibitemShut {NoStop}%
\bibitem [{\citenamefont {Liu}\ \emph {et~al.}(2019)\citenamefont {Liu},
  \citenamefont {Ke}, \citenamefont {Zhang},\ and\ \citenamefont
  {Lee}}]{Liu2019}%
  \BibitemOpen
  \bibfield  {author} {\bibinfo {author} {\bibfnamefont {W.}~\bibnamefont
  {Liu}}, \bibinfo {author} {\bibfnamefont {Y.}~\bibnamefont {Ke}}, \bibinfo
  {author} {\bibfnamefont {L.}~\bibnamefont {Zhang}}, \ and\ \bibinfo {author}
  {\bibfnamefont {C.}~\bibnamefont {Lee}},\ }\href {\doibase
  10.1103/PhysRevA.99.063614} {\bibfield  {journal} {\bibinfo  {journal} {Phys.
  Rev. A}\ }\textbf {\bibinfo {volume} {99}},\ \bibinfo {pages} {063614}
  (\bibinfo {year} {2019})}\BibitemShut {NoStop}%
\bibitem [{\citenamefont {Weitenberg}\ \emph {et~al.}(2011)\citenamefont
  {Weitenberg}, \citenamefont {Endres}, \citenamefont {Sherson}, \citenamefont
  {Cheneau}, \citenamefont {Schau{\ss}}, \citenamefont {Fukuhara},
  \citenamefont {Bloch},\ and\ \citenamefont {Kuhr}}]{Weitenberg2011}%
  \BibitemOpen
  \bibfield  {author} {\bibinfo {author} {\bibfnamefont {C.}~\bibnamefont
  {Weitenberg}}, \bibinfo {author} {\bibfnamefont {M.}~\bibnamefont {Endres}},
  \bibinfo {author} {\bibfnamefont {J.~F.}\ \bibnamefont {Sherson}}, \bibinfo
  {author} {\bibfnamefont {M.}~\bibnamefont {Cheneau}}, \bibinfo {author}
  {\bibfnamefont {P.}~\bibnamefont {Schau{\ss}}}, \bibinfo {author}
  {\bibfnamefont {T.}~\bibnamefont {Fukuhara}}, \bibinfo {author}
  {\bibfnamefont {I.}~\bibnamefont {Bloch}}, \ and\ \bibinfo {author}
  {\bibfnamefont {S.}~\bibnamefont {Kuhr}},\ }\href {\doibase
  10.1038/nature09827} {\bibfield  {journal} {\bibinfo  {journal} {Nature}\
  }\textbf {\bibinfo {volume} {471}},\ \bibinfo {pages} {319} (\bibinfo {year}
  {2011})}\BibitemShut {NoStop}%
\bibitem [{\citenamefont {Fukuhara}\ \emph {et~al.}(2013)\citenamefont
  {Fukuhara}, \citenamefont {Schau{\ss}}, \citenamefont {Endres}, \citenamefont
  {Hild}, \citenamefont {Cheneau}, \citenamefont {Bloch},\ and\ \citenamefont
  {Gross}}]{Fukuhara2013}%
  \BibitemOpen
  \bibfield  {author} {\bibinfo {author} {\bibfnamefont {T.}~\bibnamefont
  {Fukuhara}}, \bibinfo {author} {\bibfnamefont {P.}~\bibnamefont
  {Schau{\ss}}}, \bibinfo {author} {\bibfnamefont {M.}~\bibnamefont {Endres}},
  \bibinfo {author} {\bibfnamefont {S.}~\bibnamefont {Hild}}, \bibinfo {author}
  {\bibfnamefont {M.}~\bibnamefont {Cheneau}}, \bibinfo {author} {\bibfnamefont
  {I.}~\bibnamefont {Bloch}}, \ and\ \bibinfo {author} {\bibfnamefont
  {C.}~\bibnamefont {Gross}},\ }\href {\doibase 10.1038/nature12541} {\bibfield
   {journal} {\bibinfo  {journal} {Nature}\ }\textbf {\bibinfo {volume}
  {502}},\ \bibinfo {pages} {76} (\bibinfo {year} {2013})}\BibitemShut
  {NoStop}%
\bibitem [{\citenamefont {Manouchehri}\ and\ \citenamefont
  {Wang}(2014)}]{Manouchehri2014}%
  \BibitemOpen
  \bibfield  {author} {\bibinfo {author} {\bibfnamefont {K.}~\bibnamefont
  {Manouchehri}}\ and\ \bibinfo {author} {\bibfnamefont {J.}~\bibnamefont
  {Wang}},\ }\href {\doibase 10.1007/978-3-642-36014-5} {\emph {\bibinfo
  {title} {Physical Implementation of Quantum Walks}}},\ Quantum Science and
  Technology\ (\bibinfo  {publisher} {Springer},\ \bibinfo {address}
  {Netherlands},\ \bibinfo {year} {2014})\BibitemShut {NoStop}%
\bibitem [{\citenamefont {Peruzzo}\ \emph {et~al.}(2010)\citenamefont
  {Peruzzo}, \citenamefont {Lobino}, \citenamefont {Matthews}, \citenamefont
  {Matsuda}, \citenamefont {Politi}, \citenamefont {Poulios}, \citenamefont
  {Zhou}, \citenamefont {Lahini}, \citenamefont {Ismail}, \citenamefont
  {W{\"o}rhoff}, \citenamefont {Bromberg}, \citenamefont {Silberberg},
  \citenamefont {Thompson},\ and\ \citenamefont {OBrien}}]{Peruzzo2010}%
  \BibitemOpen
  \bibfield  {author} {\bibinfo {author} {\bibfnamefont {A.}~\bibnamefont
  {Peruzzo}}, \bibinfo {author} {\bibfnamefont {M.}~\bibnamefont {Lobino}},
  \bibinfo {author} {\bibfnamefont {J.~C.~F.}\ \bibnamefont {Matthews}},
  \bibinfo {author} {\bibfnamefont {N.}~\bibnamefont {Matsuda}}, \bibinfo
  {author} {\bibfnamefont {A.}~\bibnamefont {Politi}}, \bibinfo {author}
  {\bibfnamefont {K.}~\bibnamefont {Poulios}}, \bibinfo {author} {\bibfnamefont
  {X.-Q.}\ \bibnamefont {Zhou}}, \bibinfo {author} {\bibfnamefont
  {Y.}~\bibnamefont {Lahini}}, \bibinfo {author} {\bibfnamefont
  {N.}~\bibnamefont {Ismail}}, \bibinfo {author} {\bibfnamefont
  {K.}~\bibnamefont {W{\"o}rhoff}}, \bibinfo {author} {\bibfnamefont
  {Y.}~\bibnamefont {Bromberg}}, \bibinfo {author} {\bibfnamefont
  {Y.}~\bibnamefont {Silberberg}}, \bibinfo {author} {\bibfnamefont {M.~G.}\
  \bibnamefont {Thompson}}, \ and\ \bibinfo {author} {\bibfnamefont {J.~L.}\
  \bibnamefont {OBrien}},\ }\href {\doibase 10.1126/science.1193515} {\bibfield
   {journal} {\bibinfo  {journal} {Science}\ }\textbf {\bibinfo {volume}
  {329}},\ \bibinfo {pages} {1500} (\bibinfo {year} {2010})}\BibitemShut
  {NoStop}%
\bibitem [{\citenamefont {Z\"ahringer}\ \emph {et~al.}(2010)\citenamefont
  {Z\"ahringer}, \citenamefont {Kirchmair}, \citenamefont {Gerritsma},
  \citenamefont {Solano}, \citenamefont {Blatt},\ and\ \citenamefont
  {Roos}}]{Zahringer2010}%
  \BibitemOpen
  \bibfield  {author} {\bibinfo {author} {\bibfnamefont {F.}~\bibnamefont
  {Z\"ahringer}}, \bibinfo {author} {\bibfnamefont {G.}~\bibnamefont
  {Kirchmair}}, \bibinfo {author} {\bibfnamefont {R.}~\bibnamefont
  {Gerritsma}}, \bibinfo {author} {\bibfnamefont {E.}~\bibnamefont {Solano}},
  \bibinfo {author} {\bibfnamefont {R.}~\bibnamefont {Blatt}}, \ and\ \bibinfo
  {author} {\bibfnamefont {C.~F.}\ \bibnamefont {Roos}},\ }\href {\doibase
  10.1103/PhysRevLett.104.100503} {\bibfield  {journal} {\bibinfo  {journal}
  {Phys. Rev. Lett.}\ }\textbf {\bibinfo {volume} {104}},\ \bibinfo {pages}
  {100503} (\bibinfo {year} {2010})}\BibitemShut {NoStop}%
\bibitem [{\citenamefont {Karski}\ \emph {et~al.}(2009)\citenamefont {Karski},
  \citenamefont {F{\"o}rster}, \citenamefont {Choi}, \citenamefont {Steffen},
  \citenamefont {Alt}, \citenamefont {Meschede},\ and\ \citenamefont
  {Widera}}]{Karski2009}%
  \BibitemOpen
  \bibfield  {author} {\bibinfo {author} {\bibfnamefont {M.}~\bibnamefont
  {Karski}}, \bibinfo {author} {\bibfnamefont {L.}~\bibnamefont {F{\"o}rster}},
  \bibinfo {author} {\bibfnamefont {J.-M.}\ \bibnamefont {Choi}}, \bibinfo
  {author} {\bibfnamefont {A.}~\bibnamefont {Steffen}}, \bibinfo {author}
  {\bibfnamefont {W.}~\bibnamefont {Alt}}, \bibinfo {author} {\bibfnamefont
  {D.}~\bibnamefont {Meschede}}, \ and\ \bibinfo {author} {\bibfnamefont
  {A.}~\bibnamefont {Widera}},\ }\href {\doibase 10.1126/science.1174436}
  {\bibfield  {journal} {\bibinfo  {journal} {Science}\ }\textbf {\bibinfo
  {volume} {325}},\ \bibinfo {pages} {174} (\bibinfo {year}
  {2009})}\BibitemShut {NoStop}%
\bibitem [{\citenamefont {Cirac}\ and\ \citenamefont
  {Zoller}(2012)}]{Cirac2012}%
  \BibitemOpen
  \bibfield  {author} {\bibinfo {author} {\bibfnamefont {J.~I.}\ \bibnamefont
  {Cirac}}\ and\ \bibinfo {author} {\bibfnamefont {P.}~\bibnamefont {Zoller}},\
  }\href {\doibase 10.1038/nphys2275} {\bibfield  {journal} {\bibinfo
  {journal} {Nat. Phys.}\ }\textbf {\bibinfo {volume} {8}},\ \bibinfo {pages}
  {264} (\bibinfo {year} {2012})}\BibitemShut {NoStop}%
\bibitem [{\citenamefont {Bloch}\ \emph {et~al.}(2012)\citenamefont {Bloch},
  \citenamefont {Dalibard},\ and\ \citenamefont {Nascimb{\`e}ne}}]{Bloch2012}%
  \BibitemOpen
  \bibfield  {author} {\bibinfo {author} {\bibfnamefont {I.}~\bibnamefont
  {Bloch}}, \bibinfo {author} {\bibfnamefont {J.}~\bibnamefont {Dalibard}}, \
  and\ \bibinfo {author} {\bibfnamefont {S.}~\bibnamefont {Nascimb{\`e}ne}},\
  }\href {\doibase 10.1038/nphys2259} {\bibfield  {journal} {\bibinfo
  {journal} {Nat. Phys.}\ }\textbf {\bibinfo {volume} {8}},\ \bibinfo {pages}
  {267} (\bibinfo {year} {2012})}\BibitemShut {NoStop}%
\bibitem [{\citenamefont {Jaksch}\ and\ \citenamefont
  {Zoller}(2005)}]{Jaksch2005}%
  \BibitemOpen
  \bibfield  {author} {\bibinfo {author} {\bibfnamefont {D.}~\bibnamefont
  {Jaksch}}\ and\ \bibinfo {author} {\bibfnamefont {P.}~\bibnamefont
  {Zoller}},\ }\href {\doibase http://dx.doi.org/10.1016/j.aop.2004.09.010}
  {\bibfield  {journal} {\bibinfo  {journal} {Ann. Phys.}\ }\textbf {\bibinfo
  {volume} {315}},\ \bibinfo {pages} {52 } (\bibinfo {year}
  {2005})}\BibitemShut {NoStop}%
\bibitem [{\citenamefont {Bloch}\ and\ \citenamefont
  {Zoller}(2006)}]{Bloch2006}%
  \BibitemOpen
  \bibfield  {author} {\bibinfo {author} {\bibfnamefont {I.}~\bibnamefont
  {Bloch}}\ and\ \bibinfo {author} {\bibfnamefont {P.}~\bibnamefont {Zoller}},\
  }\href {http://stacks.iop.org/1367-2630/8/i=8/a=E02} {\bibfield  {journal}
  {\bibinfo  {journal} {New J. Phys.}\ }\textbf {\bibinfo {volume} {8}}
  (\bibinfo {year} {2006})}\BibitemShut {NoStop}%
\bibitem [{\citenamefont {Bloch}(2008)}]{Bloch2008}%
  \BibitemOpen
  \bibfield  {author} {\bibinfo {author} {\bibfnamefont {I.}~\bibnamefont
  {Bloch}},\ }\href {\doibase 10.1126/science.1152501} {\bibfield  {journal}
  {\bibinfo  {journal} {Science}\ }\textbf {\bibinfo {volume} {319}},\ \bibinfo
  {pages} {1202} (\bibinfo {year} {2008})}\BibitemShut {NoStop}%
\bibitem [{\citenamefont {Lewenstein}\ \emph {et~al.}(2007)\citenamefont
  {Lewenstein}, \citenamefont {Sanpera}, \citenamefont {Ahufinger},
  \citenamefont {Damski}, \citenamefont {Sen(De)},\ and\ \citenamefont
  {Sen}}]{Lewenstein2007}%
  \BibitemOpen
  \bibfield  {author} {\bibinfo {author} {\bibfnamefont {M.}~\bibnamefont
  {Lewenstein}}, \bibinfo {author} {\bibfnamefont {A.}~\bibnamefont {Sanpera}},
  \bibinfo {author} {\bibfnamefont {V.}~\bibnamefont {Ahufinger}}, \bibinfo
  {author} {\bibfnamefont {B.}~\bibnamefont {Damski}}, \bibinfo {author}
  {\bibfnamefont {A.}~\bibnamefont {Sen(De)}}, \ and\ \bibinfo {author}
  {\bibfnamefont {U.}~\bibnamefont {Sen}},\ }\href {\doibase
  10.1080/00018730701223200} {\bibfield  {journal} {\bibinfo  {journal}
  {Advances in Physics}\ }\textbf {\bibinfo {volume} {56}},\ \bibinfo {pages}
  {243} (\bibinfo {year} {2007})}\BibitemShut {NoStop}%
\bibitem [{\citenamefont {Lewenstein}\ \emph {et~al.}(2017)\citenamefont
  {Lewenstein}, \citenamefont {Sanpera},\ and\ \citenamefont
  {Ahufinger}}]{Lewenstein2017}%
  \BibitemOpen
  \bibfield  {author} {\bibinfo {author} {\bibfnamefont {M.}~\bibnamefont
  {Lewenstein}}, \bibinfo {author} {\bibfnamefont {A.}~\bibnamefont {Sanpera}},
  \ and\ \bibinfo {author} {\bibfnamefont {V.}~\bibnamefont {Ahufinger}},\
  }\href
  {https://www.oxfordscholarship.com/view/10.1093/acprof:oso/9780199573127.001.0001/acprof-9780199573127}
  {\emph {\bibinfo {title} {Ultracold Atoms in Optical Lattices: Simulating
  Quantum Many-Body Systems}}}\ (\bibinfo  {publisher} {Oxford University
  Press},\ \bibinfo {year} {2017})\BibitemShut {NoStop}%
\bibitem [{\citenamefont {Penna}\ and\ \citenamefont
  {Richaud}(2017)}]{Penna2017}%
  \BibitemOpen
  \bibfield  {author} {\bibinfo {author} {\bibfnamefont {V.}~\bibnamefont
  {Penna}}\ and\ \bibinfo {author} {\bibfnamefont {A.}~\bibnamefont
  {Richaud}},\ }\href {\doibase 10.1103/PhysRevA.96.053631} {\bibfield
  {journal} {\bibinfo  {journal} {Phys. Rev. A}\ }\textbf {\bibinfo {volume}
  {96}},\ \bibinfo {pages} {053631} (\bibinfo {year} {2017})}\BibitemShut
  {NoStop}%
\bibitem [{\citenamefont {Richaud}\ and\ \citenamefont
  {Penna}(2017)}]{Richaud2017}%
  \BibitemOpen
  \bibfield  {author} {\bibinfo {author} {\bibfnamefont {A.}~\bibnamefont
  {Richaud}}\ and\ \bibinfo {author} {\bibfnamefont {V.}~\bibnamefont
  {Penna}},\ }\href {\doibase 10.1103/PhysRevA.96.013620} {\bibfield  {journal}
  {\bibinfo  {journal} {Phys. Rev. A}\ }\textbf {\bibinfo {volume} {96}},\
  \bibinfo {pages} {013620} (\bibinfo {year} {2017})}\BibitemShut {NoStop}%
\bibitem [{\citenamefont {Keiler}\ and\ \citenamefont
  {Schmelcher}(2018)}]{Keiler2018}%
  \BibitemOpen
  \bibfield  {author} {\bibinfo {author} {\bibfnamefont {K.}~\bibnamefont
  {Keiler}}\ and\ \bibinfo {author} {\bibfnamefont {P.}~\bibnamefont
  {Schmelcher}},\ }\href {\doibase 10.1088/1367-2630/aae98f} {\bibfield
  {journal} {\bibinfo  {journal} {New Journal of Physics}\ }\textbf {\bibinfo
  {volume} {20}},\ \bibinfo {pages} {103042} (\bibinfo {year}
  {2018})}\BibitemShut {NoStop}%
\bibitem [{\citenamefont {Keiler}\ and\ \citenamefont
  {Schmelcher}(2019)}]{Keiler2019}%
  \BibitemOpen
  \bibfield  {author} {\bibinfo {author} {\bibfnamefont {K.}~\bibnamefont
  {Keiler}}\ and\ \bibinfo {author} {\bibfnamefont {P.}~\bibnamefont
  {Schmelcher}},\ }\href {\doibase 10.1103/PhysRevA.100.043616} {\bibfield
  {journal} {\bibinfo  {journal} {Phys. Rev. A}\ }\textbf {\bibinfo {volume}
  {100}},\ \bibinfo {pages} {043616} (\bibinfo {year} {2019})}\BibitemShut
  {NoStop}%
\bibitem [{\citenamefont {Bromberg}\ \emph {et~al.}(2009)\citenamefont
  {Bromberg}, \citenamefont {Lahini}, \citenamefont {Morandotti},\ and\
  \citenamefont {Silberberg}}]{Bromberg2009}%
  \BibitemOpen
  \bibfield  {author} {\bibinfo {author} {\bibfnamefont {Y.}~\bibnamefont
  {Bromberg}}, \bibinfo {author} {\bibfnamefont {Y.}~\bibnamefont {Lahini}},
  \bibinfo {author} {\bibfnamefont {R.}~\bibnamefont {Morandotti}}, \ and\
  \bibinfo {author} {\bibfnamefont {Y.}~\bibnamefont {Silberberg}},\ }\href
  {\doibase 10.1103/PhysRevLett.102.253904} {\bibfield  {journal} {\bibinfo
  {journal} {Phys. Rev. Lett.}\ }\textbf {\bibinfo {volume} {102}},\ \bibinfo
  {pages} {253904} (\bibinfo {year} {2009})}\BibitemShut {NoStop}%
\bibitem [{\citenamefont {Ahlbrecht}\ \emph {et~al.}(2012)\citenamefont
  {Ahlbrecht}, \citenamefont {Alberti}, \citenamefont {Meschede}, \citenamefont
  {Scholz}, \citenamefont {Werner},\ and\ \citenamefont
  {Werner}}]{Ahlbrecht2012}%
  \BibitemOpen
  \bibfield  {author} {\bibinfo {author} {\bibfnamefont {A.}~\bibnamefont
  {Ahlbrecht}}, \bibinfo {author} {\bibfnamefont {A.}~\bibnamefont {Alberti}},
  \bibinfo {author} {\bibfnamefont {D.}~\bibnamefont {Meschede}}, \bibinfo
  {author} {\bibfnamefont {V.~B.}\ \bibnamefont {Scholz}}, \bibinfo {author}
  {\bibfnamefont {A.~H.}\ \bibnamefont {Werner}}, \ and\ \bibinfo {author}
  {\bibfnamefont {R.~F.}\ \bibnamefont {Werner}},\ }\href {\doibase
  10.1088/1367-2630/14/7/073050} {\bibfield  {journal} {\bibinfo  {journal}
  {New Journal of Physics}\ }\textbf {\bibinfo {volume} {14}},\ \bibinfo
  {pages} {073050} (\bibinfo {year} {2012})}\BibitemShut {NoStop}%
\bibitem [{\citenamefont {Qin}\ \emph {et~al.}(2014)\citenamefont {Qin},
  \citenamefont {Ke}, \citenamefont {Guan}, \citenamefont {Li}, \citenamefont
  {Andrei},\ and\ \citenamefont {Lee}}]{Qin2014}%
  \BibitemOpen
  \bibfield  {author} {\bibinfo {author} {\bibfnamefont {X.}~\bibnamefont
  {Qin}}, \bibinfo {author} {\bibfnamefont {Y.}~\bibnamefont {Ke}}, \bibinfo
  {author} {\bibfnamefont {X.}~\bibnamefont {Guan}}, \bibinfo {author}
  {\bibfnamefont {Z.}~\bibnamefont {Li}}, \bibinfo {author} {\bibfnamefont
  {N.}~\bibnamefont {Andrei}}, \ and\ \bibinfo {author} {\bibfnamefont
  {C.}~\bibnamefont {Lee}},\ }\href {\doibase 10.1103/PhysRevA.90.062301}
  {\bibfield  {journal} {\bibinfo  {journal} {Phys. Rev. A}\ }\textbf {\bibinfo
  {volume} {90}},\ \bibinfo {pages} {062301} (\bibinfo {year}
  {2014})}\BibitemShut {NoStop}%
\bibitem [{\citenamefont {Childs}\ \emph {et~al.}(2013)\citenamefont {Childs},
  \citenamefont {Gosset},\ and\ \citenamefont {Webb}}]{Childs2013}%
  \BibitemOpen
  \bibfield  {author} {\bibinfo {author} {\bibfnamefont {A.~M.}\ \bibnamefont
  {Childs}}, \bibinfo {author} {\bibfnamefont {D.}~\bibnamefont {Gosset}}, \
  and\ \bibinfo {author} {\bibfnamefont {Z.}~\bibnamefont {Webb}},\ }\href
  {\doibase 10.1126/science.1229957} {\bibfield  {journal} {\bibinfo  {journal}
  {Science}\ }\textbf {\bibinfo {volume} {339}},\ \bibinfo {pages} {791}
  (\bibinfo {year} {2013})}\BibitemShut {NoStop}%
\bibitem [{\citenamefont {Bloch}(1929)}]{1929BlochZeit}%
  \BibitemOpen
  \bibfield  {author} {\bibinfo {author} {\bibfnamefont {F.}~\bibnamefont
  {Bloch}},\ }\href {\doibase 10.1007/BF01339455} {\bibfield  {journal}
  {\bibinfo  {journal} {Zeitschrift f{\"u}r Physik}\ }\textbf {\bibinfo
  {volume} {52}},\ \bibinfo {pages} {555} (\bibinfo {year} {1929})}\BibitemShut
  {NoStop}%
\bibitem [{\citenamefont {Leo}\ \emph {et~al.}(1992)\citenamefont {Leo},
  \citenamefont {Bolivar}, \citenamefont {Brüggemann}, \citenamefont
  {Schwedler},\ and\ \citenamefont {Köhler}}]{1992LeoSolidStateCom}%
  \BibitemOpen
  \bibfield  {author} {\bibinfo {author} {\bibfnamefont {K.}~\bibnamefont
  {Leo}}, \bibinfo {author} {\bibfnamefont {P.~H.}\ \bibnamefont {Bolivar}},
  \bibinfo {author} {\bibfnamefont {F.}~\bibnamefont {Brüggemann}}, \bibinfo
  {author} {\bibfnamefont {R.}~\bibnamefont {Schwedler}}, \ and\ \bibinfo
  {author} {\bibfnamefont {K.}~\bibnamefont {Köhler}},\ }\href {\doibase
  https://doi.org/10.1016/0038-1098(92)90798-E} {\bibfield  {journal} {\bibinfo
   {journal} {Solid State Communications}\ }\textbf {\bibinfo {volume} {84}},\
  \bibinfo {pages} {943 } (\bibinfo {year} {1992})}\BibitemShut {NoStop}%
\bibitem [{\citenamefont {Ben~Dahan}\ \emph {et~al.}(1996)\citenamefont
  {Ben~Dahan}, \citenamefont {Peik}, \citenamefont {Reichel}, \citenamefont
  {Castin},\ and\ \citenamefont {Salomon}}]{Dahan1996}%
  \BibitemOpen
  \bibfield  {author} {\bibinfo {author} {\bibfnamefont {M.}~\bibnamefont
  {Ben~Dahan}}, \bibinfo {author} {\bibfnamefont {E.}~\bibnamefont {Peik}},
  \bibinfo {author} {\bibfnamefont {J.}~\bibnamefont {Reichel}}, \bibinfo
  {author} {\bibfnamefont {Y.}~\bibnamefont {Castin}}, \ and\ \bibinfo {author}
  {\bibfnamefont {C.}~\bibnamefont {Salomon}},\ }\href {\doibase
  10.1103/PhysRevLett.76.4508} {\bibfield  {journal} {\bibinfo  {journal}
  {Phys. Rev. Lett.}\ }\textbf {\bibinfo {volume} {76}},\ \bibinfo {pages}
  {4508} (\bibinfo {year} {1996})}\BibitemShut {NoStop}%
\bibitem [{\citenamefont {Wilkinson}\ \emph {et~al.}(1996)\citenamefont
  {Wilkinson}, \citenamefont {Bharucha}, \citenamefont {Madison}, \citenamefont
  {Niu},\ and\ \citenamefont {Raizen}}]{Wilkinson1996}%
  \BibitemOpen
  \bibfield  {author} {\bibinfo {author} {\bibfnamefont {S.~R.}\ \bibnamefont
  {Wilkinson}}, \bibinfo {author} {\bibfnamefont {C.~F.}\ \bibnamefont
  {Bharucha}}, \bibinfo {author} {\bibfnamefont {K.~W.}\ \bibnamefont
  {Madison}}, \bibinfo {author} {\bibfnamefont {Q.}~\bibnamefont {Niu}}, \ and\
  \bibinfo {author} {\bibfnamefont {M.~G.}\ \bibnamefont {Raizen}},\ }\href
  {\doibase 10.1103/PhysRevLett.76.4512} {\bibfield  {journal} {\bibinfo
  {journal} {Phys. Rev. Lett.}\ }\textbf {\bibinfo {volume} {76}},\ \bibinfo
  {pages} {4512} (\bibinfo {year} {1996})}\BibitemShut {NoStop}%
\bibitem [{\citenamefont {Ferrari}\ \emph {et~al.}(2006)\citenamefont
  {Ferrari}, \citenamefont {Poli}, \citenamefont {Sorrentino},\ and\
  \citenamefont {Tino}}]{Ferrari2006}%
  \BibitemOpen
  \bibfield  {author} {\bibinfo {author} {\bibfnamefont {G.}~\bibnamefont
  {Ferrari}}, \bibinfo {author} {\bibfnamefont {N.}~\bibnamefont {Poli}},
  \bibinfo {author} {\bibfnamefont {F.}~\bibnamefont {Sorrentino}}, \ and\
  \bibinfo {author} {\bibfnamefont {G.~M.}\ \bibnamefont {Tino}},\ }\href
  {\doibase 10.1103/PhysRevLett.97.060402} {\bibfield  {journal} {\bibinfo
  {journal} {Phys. Rev. Lett.}\ }\textbf {\bibinfo {volume} {97}},\ \bibinfo
  {pages} {060402} (\bibinfo {year} {2006})}\BibitemShut {NoStop}%
\bibitem [{\citenamefont {Poli}\ \emph {et~al.}(2011)\citenamefont {Poli},
  \citenamefont {Wang}, \citenamefont {Tarallo}, \citenamefont {Alberti},
  \citenamefont {Prevedelli},\ and\ \citenamefont {Tino}}]{Poli2011}%
  \BibitemOpen
  \bibfield  {author} {\bibinfo {author} {\bibfnamefont {N.}~\bibnamefont
  {Poli}}, \bibinfo {author} {\bibfnamefont {F.-Y.}\ \bibnamefont {Wang}},
  \bibinfo {author} {\bibfnamefont {M.~G.}\ \bibnamefont {Tarallo}}, \bibinfo
  {author} {\bibfnamefont {A.}~\bibnamefont {Alberti}}, \bibinfo {author}
  {\bibfnamefont {M.}~\bibnamefont {Prevedelli}}, \ and\ \bibinfo {author}
  {\bibfnamefont {G.~M.}\ \bibnamefont {Tino}},\ }\href {\doibase
  10.1103/PhysRevLett.106.038501} {\bibfield  {journal} {\bibinfo  {journal}
  {Phys. Rev. Lett.}\ }\textbf {\bibinfo {volume} {106}},\ \bibinfo {pages}
  {038501} (\bibinfo {year} {2011})}\BibitemShut {NoStop}%
\bibitem [{\citenamefont {Gustavsson}\ \emph {et~al.}(2008)\citenamefont
  {Gustavsson}, \citenamefont {Haller}, \citenamefont {Mark}, \citenamefont
  {Danzl}, \citenamefont {Rojas-Kopeinig},\ and\ \citenamefont
  {N\"agerl}}]{Gustavsson2008}%
  \BibitemOpen
  \bibfield  {author} {\bibinfo {author} {\bibfnamefont {M.}~\bibnamefont
  {Gustavsson}}, \bibinfo {author} {\bibfnamefont {E.}~\bibnamefont {Haller}},
  \bibinfo {author} {\bibfnamefont {M.~J.}\ \bibnamefont {Mark}}, \bibinfo
  {author} {\bibfnamefont {J.~G.}\ \bibnamefont {Danzl}}, \bibinfo {author}
  {\bibfnamefont {G.}~\bibnamefont {Rojas-Kopeinig}}, \ and\ \bibinfo {author}
  {\bibfnamefont {H.-C.}\ \bibnamefont {N\"agerl}},\ }\href {\doibase
  10.1103/PhysRevLett.100.080404} {\bibfield  {journal} {\bibinfo  {journal}
  {Phys. Rev. Lett.}\ }\textbf {\bibinfo {volume} {100}},\ \bibinfo {pages}
  {080404} (\bibinfo {year} {2008})}\BibitemShut {NoStop}%
\bibitem [{\citenamefont {Fattori}\ \emph {et~al.}(2008)\citenamefont
  {Fattori}, \citenamefont {D'Errico}, \citenamefont {Roati}, \citenamefont
  {Zaccanti}, \citenamefont {Jona-Lasinio}, \citenamefont {Modugno},
  \citenamefont {Inguscio},\ and\ \citenamefont {Modugno}}]{Fattori2008}%
  \BibitemOpen
  \bibfield  {author} {\bibinfo {author} {\bibfnamefont {M.}~\bibnamefont
  {Fattori}}, \bibinfo {author} {\bibfnamefont {C.}~\bibnamefont {D'Errico}},
  \bibinfo {author} {\bibfnamefont {G.}~\bibnamefont {Roati}}, \bibinfo
  {author} {\bibfnamefont {M.}~\bibnamefont {Zaccanti}}, \bibinfo {author}
  {\bibfnamefont {M.}~\bibnamefont {Jona-Lasinio}}, \bibinfo {author}
  {\bibfnamefont {M.}~\bibnamefont {Modugno}}, \bibinfo {author} {\bibfnamefont
  {M.}~\bibnamefont {Inguscio}}, \ and\ \bibinfo {author} {\bibfnamefont
  {G.}~\bibnamefont {Modugno}},\ }\href {\doibase
  10.1103/PhysRevLett.100.080405} {\bibfield  {journal} {\bibinfo  {journal}
  {Phys. Rev. Lett.}\ }\textbf {\bibinfo {volume} {100}},\ \bibinfo {pages}
  {080405} (\bibinfo {year} {2008})}\BibitemShut {NoStop}%
\bibitem [{\citenamefont {Alberti}\ \emph {et~al.}(2009)\citenamefont
  {Alberti}, \citenamefont {Ivanov}, \citenamefont {Tino},\ and\ \citenamefont
  {Ferrari}}]{Alberti2009}%
  \BibitemOpen
  \bibfield  {author} {\bibinfo {author} {\bibfnamefont {A.}~\bibnamefont
  {Alberti}}, \bibinfo {author} {\bibfnamefont {V.~V.}\ \bibnamefont {Ivanov}},
  \bibinfo {author} {\bibfnamefont {G.~M.}\ \bibnamefont {Tino}}, \ and\
  \bibinfo {author} {\bibfnamefont {G.}~\bibnamefont {Ferrari}},\ }\href
  {\doibase 10.1038/nphys1310} {\bibfield  {journal} {\bibinfo  {journal}
  {Nature Physics}\ }\textbf {\bibinfo {volume} {5}},\ \bibinfo {pages} {547}
  (\bibinfo {year} {2009})}\BibitemShut {NoStop}%
\bibitem [{\citenamefont {Haller}\ \emph {et~al.}(2010)\citenamefont {Haller},
  \citenamefont {Hart}, \citenamefont {Mark}, \citenamefont {Danzl},
  \citenamefont {Reichs\"ollner},\ and\ \citenamefont {N\"agerl}}]{Haller2010}%
  \BibitemOpen
  \bibfield  {author} {\bibinfo {author} {\bibfnamefont {E.}~\bibnamefont
  {Haller}}, \bibinfo {author} {\bibfnamefont {R.}~\bibnamefont {Hart}},
  \bibinfo {author} {\bibfnamefont {M.~J.}\ \bibnamefont {Mark}}, \bibinfo
  {author} {\bibfnamefont {J.~G.}\ \bibnamefont {Danzl}}, \bibinfo {author}
  {\bibfnamefont {L.}~\bibnamefont {Reichs\"ollner}}, \ and\ \bibinfo {author}
  {\bibfnamefont {H.-C.}\ \bibnamefont {N\"agerl}},\ }\href {\doibase
  10.1103/PhysRevLett.104.200403} {\bibfield  {journal} {\bibinfo  {journal}
  {Phys. Rev. Lett.}\ }\textbf {\bibinfo {volume} {104}},\ \bibinfo {pages}
  {200403} (\bibinfo {year} {2010})}\BibitemShut {NoStop}%
\bibitem [{\citenamefont {Zhang}\ and\ \citenamefont {Liu}(2010)}]{Zhang2010}%
  \BibitemOpen
  \bibfield  {author} {\bibinfo {author} {\bibfnamefont {J.~M.}\ \bibnamefont
  {Zhang}}\ and\ \bibinfo {author} {\bibfnamefont {W.~M.}\ \bibnamefont
  {Liu}},\ }\href {\doibase 10.1103/PhysRevA.82.025602} {\bibfield  {journal}
  {\bibinfo  {journal} {Phys. Rev. A}\ }\textbf {\bibinfo {volume} {82}},\
  \bibinfo {pages} {025602} (\bibinfo {year} {2010})}\BibitemShut {NoStop}%
\bibitem [{\citenamefont {Lyssenko}\ \emph {et~al.}(1997)\citenamefont
  {Lyssenko}, \citenamefont {Valu\ifmmode~\check{s}\else \v{s}\fi{}is},
  \citenamefont {L\"oser}, \citenamefont {Hasche}, \citenamefont {Leo},
  \citenamefont {Dignam},\ and\ \citenamefont {K\"ohler}}]{Lyssenko1997}%
  \BibitemOpen
  \bibfield  {author} {\bibinfo {author} {\bibfnamefont {V.~G.}\ \bibnamefont
  {Lyssenko}}, \bibinfo {author} {\bibfnamefont {G.}~\bibnamefont
  {Valu\ifmmode~\check{s}\else \v{s}\fi{}is}}, \bibinfo {author} {\bibfnamefont
  {F.}~\bibnamefont {L\"oser}}, \bibinfo {author} {\bibfnamefont
  {T.}~\bibnamefont {Hasche}}, \bibinfo {author} {\bibfnamefont
  {K.}~\bibnamefont {Leo}}, \bibinfo {author} {\bibfnamefont {M.~M.}\
  \bibnamefont {Dignam}}, \ and\ \bibinfo {author} {\bibfnamefont
  {K.}~\bibnamefont {K\"ohler}},\ }\href {\doibase 10.1103/PhysRevLett.79.301}
  {\bibfield  {journal} {\bibinfo  {journal} {Phys. Rev. Lett.}\ }\textbf
  {\bibinfo {volume} {79}},\ \bibinfo {pages} {301} (\bibinfo {year}
  {1997})}\BibitemShut {NoStop}%
\bibitem [{\citenamefont {Hartmann}\ \emph {et~al.}(2004)\citenamefont
  {Hartmann}, \citenamefont {Keck}, \citenamefont {Korsch},\ and\ \citenamefont
  {Mossmann}}]{Hartmann2004}%
  \BibitemOpen
  \bibfield  {author} {\bibinfo {author} {\bibfnamefont {T.}~\bibnamefont
  {Hartmann}}, \bibinfo {author} {\bibfnamefont {F.}~\bibnamefont {Keck}},
  \bibinfo {author} {\bibfnamefont {H.~J.}\ \bibnamefont {Korsch}}, \ and\
  \bibinfo {author} {\bibfnamefont {S.}~\bibnamefont {Mossmann}},\ }\href
  {\doibase 10.1088/1367-2630/6/1/002} {\bibfield  {journal} {\bibinfo
  {journal} {New Journal of Physics}\ }\textbf {\bibinfo {volume} {6}},\
  \bibinfo {pages} {2} (\bibinfo {year} {2004})}\BibitemShut {NoStop}%
\bibitem [{\citenamefont {Dias}\ \emph {et~al.}(2007)\citenamefont {Dias},
  \citenamefont {Nascimento}, \citenamefont {Lyra},\ and\ \citenamefont
  {de~Moura}}]{Dias2007}%
  \BibitemOpen
  \bibfield  {author} {\bibinfo {author} {\bibfnamefont {W.~S.}\ \bibnamefont
  {Dias}}, \bibinfo {author} {\bibfnamefont {E.~M.}\ \bibnamefont
  {Nascimento}}, \bibinfo {author} {\bibfnamefont {M.~L.}\ \bibnamefont
  {Lyra}}, \ and\ \bibinfo {author} {\bibfnamefont {F.~A. B.~F.}\ \bibnamefont
  {de~Moura}},\ }\href {\doibase 10.1103/PhysRevB.76.155124} {\bibfield
  {journal} {\bibinfo  {journal} {Phys. Rev. B}\ }\textbf {\bibinfo {volume}
  {76}},\ \bibinfo {pages} {155124} (\bibinfo {year} {2007})}\BibitemShut
  {NoStop}%
\bibitem [{\citenamefont {Khomeriki}\ \emph {et~al.}(2010)\citenamefont
  {Khomeriki}, \citenamefont {Krimer}, \citenamefont {Haque},\ and\
  \citenamefont {Flach}}]{Khomeriki2010}%
  \BibitemOpen
  \bibfield  {author} {\bibinfo {author} {\bibfnamefont {R.}~\bibnamefont
  {Khomeriki}}, \bibinfo {author} {\bibfnamefont {D.~O.}\ \bibnamefont
  {Krimer}}, \bibinfo {author} {\bibfnamefont {M.}~\bibnamefont {Haque}}, \
  and\ \bibinfo {author} {\bibfnamefont {S.}~\bibnamefont {Flach}},\ }\href
  {\doibase 10.1103/PhysRevA.81.065601} {\bibfield  {journal} {\bibinfo
  {journal} {Phys. Rev. A}\ }\textbf {\bibinfo {volume} {81}},\ \bibinfo
  {pages} {065601} (\bibinfo {year} {2010})}\BibitemShut {NoStop}%
\bibitem [{\citenamefont {Sachdev}\ \emph {et~al.}(2002)\citenamefont
  {Sachdev}, \citenamefont {Sengupta},\ and\ \citenamefont
  {Girvin}}]{Sachdev2002}%
  \BibitemOpen
  \bibfield  {author} {\bibinfo {author} {\bibfnamefont {S.}~\bibnamefont
  {Sachdev}}, \bibinfo {author} {\bibfnamefont {K.}~\bibnamefont {Sengupta}}, \
  and\ \bibinfo {author} {\bibfnamefont {S.~M.}\ \bibnamefont {Girvin}},\
  }\href {\doibase 10.1103/PhysRevB.66.075128} {\bibfield  {journal} {\bibinfo
  {journal} {Phys. Rev. B}\ }\textbf {\bibinfo {volume} {66}},\ \bibinfo
  {pages} {075128} (\bibinfo {year} {2002})}\BibitemShut {NoStop}%
\bibitem [{\citenamefont {Meinert}\ \emph {et~al.}(2014)\citenamefont
  {Meinert}, \citenamefont {Mark}, \citenamefont {Kirilov}, \citenamefont
  {Lauber}, \citenamefont {Weinmann}, \citenamefont {Gr\"obner},\ and\
  \citenamefont {N\"agerl}}]{Meinert2014}%
  \BibitemOpen
  \bibfield  {author} {\bibinfo {author} {\bibfnamefont {F.}~\bibnamefont
  {Meinert}}, \bibinfo {author} {\bibfnamefont {M.~J.}\ \bibnamefont {Mark}},
  \bibinfo {author} {\bibfnamefont {E.}~\bibnamefont {Kirilov}}, \bibinfo
  {author} {\bibfnamefont {K.}~\bibnamefont {Lauber}}, \bibinfo {author}
  {\bibfnamefont {P.}~\bibnamefont {Weinmann}}, \bibinfo {author}
  {\bibfnamefont {M.}~\bibnamefont {Gr\"obner}}, \ and\ \bibinfo {author}
  {\bibfnamefont {H.-C.}\ \bibnamefont {N\"agerl}},\ }\href {\doibase
  10.1103/PhysRevLett.112.193003} {\bibfield  {journal} {\bibinfo  {journal}
  {Phys. Rev. Lett.}\ }\textbf {\bibinfo {volume} {112}},\ \bibinfo {pages}
  {193003} (\bibinfo {year} {2014})}\BibitemShut {NoStop}%
\bibitem [{\citenamefont {Preiss}\ \emph {et~al.}(2015)\citenamefont {Preiss},
  \citenamefont {Ma}, \citenamefont {Tai}, \citenamefont {Lukin}, \citenamefont
  {Rispoli}, \citenamefont {Zupancic}, \citenamefont {Lahini}, \citenamefont
  {Islam},\ and\ \citenamefont {Greiner}}]{Preiss2015}%
  \BibitemOpen
  \bibfield  {author} {\bibinfo {author} {\bibfnamefont {P.~M.}\ \bibnamefont
  {Preiss}}, \bibinfo {author} {\bibfnamefont {R.}~\bibnamefont {Ma}}, \bibinfo
  {author} {\bibfnamefont {M.~E.}\ \bibnamefont {Tai}}, \bibinfo {author}
  {\bibfnamefont {A.}~\bibnamefont {Lukin}}, \bibinfo {author} {\bibfnamefont
  {M.}~\bibnamefont {Rispoli}}, \bibinfo {author} {\bibfnamefont
  {P.}~\bibnamefont {Zupancic}}, \bibinfo {author} {\bibfnamefont
  {Y.}~\bibnamefont {Lahini}}, \bibinfo {author} {\bibfnamefont
  {R.}~\bibnamefont {Islam}}, \ and\ \bibinfo {author} {\bibfnamefont
  {M.}~\bibnamefont {Greiner}},\ }\href {\doibase 10.1126/science.1260364}
  {\bibfield  {journal} {\bibinfo  {journal} {Science}\ }\textbf {\bibinfo
  {volume} {347}},\ \bibinfo {pages} {1229} (\bibinfo {year}
  {2015})}\BibitemShut {NoStop}%
\bibitem [{\citenamefont {Wiater}\ \emph {et~al.}(2017)\citenamefont {Wiater},
  \citenamefont {Sowi\'{n}ski},\ and\ \citenamefont {Zakrzewski}}]{Wiater2017}%
  \BibitemOpen
  \bibfield  {author} {\bibinfo {author} {\bibfnamefont {D.}~\bibnamefont
  {Wiater}}, \bibinfo {author} {\bibfnamefont {T.}~\bibnamefont
  {Sowi\'{n}ski}}, \ and\ \bibinfo {author} {\bibfnamefont {J.}~\bibnamefont
  {Zakrzewski}},\ }\href {\doibase 10.1103/PhysRevA.96.043629} {\bibfield
  {journal} {\bibinfo  {journal} {Phys. Rev. A}\ }\textbf {\bibinfo {volume}
  {96}},\ \bibinfo {pages} {043629} (\bibinfo {year} {2017})}\BibitemShut
  {NoStop}%
\bibitem [{\citenamefont {Mondal}\ and\ \citenamefont
  {Mishra}(2020)}]{Mondal2020}%
  \BibitemOpen
  \bibfield  {author} {\bibinfo {author} {\bibfnamefont {S.}~\bibnamefont
  {Mondal}}\ and\ \bibinfo {author} {\bibfnamefont {T.}~\bibnamefont
  {Mishra}},\ }\href {\doibase 10.1103/PhysRevA.101.052341} {\bibfield
  {journal} {\bibinfo  {journal} {Phys. Rev. A}\ }\textbf {\bibinfo {volume}
  {101}},\ \bibinfo {pages} {052341} (\bibinfo {year} {2020})}\BibitemShut
  {NoStop}%
\bibitem [{\citenamefont {Alonso-Lobo}\ \emph {et~al.}(2018)\citenamefont
  {Alonso-Lobo}, \citenamefont {Mart{\'{\i}}nez-Quesada}, \citenamefont
  {Hinarejos}, \citenamefont {de~Valc{\'{a}}rcel},\ and\ \citenamefont
  {Rold{\'{a}}n}}]{Alonso_Lobo2018}%
  \BibitemOpen
  \bibfield  {author} {\bibinfo {author} {\bibfnamefont {C.}~\bibnamefont
  {Alonso-Lobo}}, \bibinfo {author} {\bibfnamefont {M.}~\bibnamefont
  {Mart{\'{\i}}nez-Quesada}}, \bibinfo {author} {\bibfnamefont
  {M.}~\bibnamefont {Hinarejos}}, \bibinfo {author} {\bibfnamefont {G.~J.}\
  \bibnamefont {de~Valc{\'{a}}rcel}}, \ and\ \bibinfo {author} {\bibfnamefont
  {E.}~\bibnamefont {Rold{\'{a}}n}},\ }\href {\doibase
  10.1088/1751-8121/aadfa2} {\bibfield  {journal} {\bibinfo  {journal} {Journal
  of Physics A: Mathematical and Theoretical}\ }\textbf {\bibinfo {volume}
  {51}},\ \bibinfo {pages} {455301} (\bibinfo {year} {2018})}\BibitemShut
  {NoStop}%
\bibitem [{\citenamefont {{H}ubbard}(1963)}]{Hubbard1963}%
  \BibitemOpen
  \bibfield  {author} {\bibinfo {author} {\bibfnamefont {J.}~\bibnamefont
  {{H}ubbard}},\ }\href {http://www.jstor.org/stable/2414761} {\bibfield
  {journal} {\bibinfo  {journal} {Proceedings of the Royal Society of London.
  Series A, Mathematical and Physical Sciences}\ }\textbf {\bibinfo {volume}
  {276}},\ \bibinfo {pages} {238} (\bibinfo {year} {1963})}\BibitemShut
  {NoStop}%
\bibitem [{\citenamefont {Jaksch}\ \emph {et~al.}(1998)\citenamefont {Jaksch},
  \citenamefont {Bruder}, \citenamefont {Cirac}, \citenamefont {Gardiner},\
  and\ \citenamefont {Zoller}}]{Jaksch1998}%
  \BibitemOpen
  \bibfield  {author} {\bibinfo {author} {\bibfnamefont {D.}~\bibnamefont
  {Jaksch}}, \bibinfo {author} {\bibfnamefont {C.}~\bibnamefont {Bruder}},
  \bibinfo {author} {\bibfnamefont {J.~I.}\ \bibnamefont {Cirac}}, \bibinfo
  {author} {\bibfnamefont {C.~W.}\ \bibnamefont {Gardiner}}, \ and\ \bibinfo
  {author} {\bibfnamefont {P.}~\bibnamefont {Zoller}},\ }\href {\doibase
  10.1103/PhysRevLett.81.3108} {\bibfield  {journal} {\bibinfo  {journal}
  {Phys. Rev. Lett.}\ }\textbf {\bibinfo {volume} {81}},\ \bibinfo {pages}
  {3108} (\bibinfo {year} {1998})}\BibitemShut {NoStop}%
\bibitem [{\citenamefont {Buchleitner}\ and\ \citenamefont
  {Kolovsky}(2003)}]{Buchleitner2003}%
  \BibitemOpen
  \bibfield  {author} {\bibinfo {author} {\bibfnamefont {A.}~\bibnamefont
  {Buchleitner}}\ and\ \bibinfo {author} {\bibfnamefont {A.~R.}\ \bibnamefont
  {Kolovsky}},\ }\href {\doibase 10.1103/PhysRevLett.91.253002} {\bibfield
  {journal} {\bibinfo  {journal} {Phys. Rev. Lett.}\ }\textbf {\bibinfo
  {volume} {91}},\ \bibinfo {pages} {253002} (\bibinfo {year}
  {2003})}\BibitemShut {NoStop}%
\bibitem [{\citenamefont {Wildermuth}\ \emph {et~al.}(2005)\citenamefont
  {Wildermuth}, \citenamefont {Hofferberth}, \citenamefont {Lesanovsky},
  \citenamefont {Haller}, \citenamefont {Andersson}, \citenamefont {Groth},
  \citenamefont {Bar-Joseph}, \citenamefont {Kr{\"u}ger},\ and\ \citenamefont
  {Schmiedmayer}}]{Wildermuth2005}%
  \BibitemOpen
  \bibfield  {author} {\bibinfo {author} {\bibfnamefont {S.}~\bibnamefont
  {Wildermuth}}, \bibinfo {author} {\bibfnamefont {S.}~\bibnamefont
  {Hofferberth}}, \bibinfo {author} {\bibfnamefont {I.}~\bibnamefont
  {Lesanovsky}}, \bibinfo {author} {\bibfnamefont {E.}~\bibnamefont {Haller}},
  \bibinfo {author} {\bibfnamefont {L.~M.}\ \bibnamefont {Andersson}}, \bibinfo
  {author} {\bibfnamefont {S.}~\bibnamefont {Groth}}, \bibinfo {author}
  {\bibfnamefont {I.}~\bibnamefont {Bar-Joseph}}, \bibinfo {author}
  {\bibfnamefont {P.}~\bibnamefont {Kr{\"u}ger}}, \ and\ \bibinfo {author}
  {\bibfnamefont {J.}~\bibnamefont {Schmiedmayer}},\ }\href {\doibase
  10.1038/435440a} {\bibfield  {journal} {\bibinfo  {journal} {Nature}\
  }\textbf {\bibinfo {volume} {435}},\ \bibinfo {pages} {440} (\bibinfo {year}
  {2005})}\BibitemShut {NoStop}%
\end{thebibliography}%

%%%%%%%%%%%%%%%%%%%%%%%%%%%%%%%%%%%%%%%%%%%%%%%%%%%%
\end{document}